\documentclass[manuscript,screen,nonacm]{acmart}

\usepackage{array}
\usepackage{tabularx}
\usepackage{listings}
\usepackage{subcaption}
\lstdefinestyle{systemprompt}{
  basicstyle=\ttfamily\small,
  columns=fullflexible,
  keepspaces=true,
  breaklines=true,
  breakatwhitespace=true,
  showstringspaces=false,
  frame=single,
  framerule=0.3pt,
  framesep=6pt,
  xleftmargin=7pt,
  xrightmargin=7pt,
  aboveskip=0.5\baselineskip,
  belowskip=0.5\baselineskip,
  literate={“}{{\textquotedblleft}}1 {”}{{\textquotedblright}}1
}

\setcopyright{none}
\acmDOI{}

\acmConference[CHI '27]{ACM CHI Conference on Human Factors in Computing Systems}{May 10--14, 2027}{Pittsburgh, PA, USA}
\acmYear{2027}
\copyrightyear{2027}

\newcolumntype{Y}{>{\raggedright\arraybackslash}X}
\newcolumntype{L}[1]{>{\raggedright\arraybackslash}p{#1}}

\newcommand{\mealstimulus}[3][0 0 0 0]{%
  \begin{minipage}[t]{0.188\textwidth}
    \centering
    \includegraphics[trim=#1,clip,width=\linewidth,height=0.75\linewidth,keepaspectratio]{figures/#2}\\[-1mm]
    {\scriptsize #3}
  \end{minipage}%
}

\begin{document}

\title{Toward Low-Burden Image-Based Dietary Assessment Through Human–AI Collaborative Correction}

\author{Darren Liu}
\affiliation{%
  \department{Department of Computer Science and Center for Data Science}
  \institution{Emory University}
  \city{Atlanta}
  \state{Georgia}
  \country{USA}
}
\email{darren.liu@emory.edu}

\author{Andrew Lu}
\affiliation{%
  \department{Center for Data Science and Nell Hodgson Woodruff School of Nursing}
  \institution{Emory University}
  \city{Atlanta}
  \state{Georgia}
  \country{USA}
}
\email{andrew.lu2@emory.edu}

\author{Zhiyuan ``Ryan'' Wang}
\affiliation{%
  \department{Department of Systems and Information Engineering}
  \institution{University of Virginia}
  \city{Charlottesville}
  \state{Virginia}
  \country{USA}
}
\email{vmf9pr@virginia.edu}

\author{Hanqi Luo}
\affiliation{%
  \department{Hubert Department of Global Health, Rollins School of Public Health}
  \institution{Emory University}
  \city{Atlanta}
  \state{Georgia}
  \country{USA}
}
\email{hluo30@emory.edu}

\author{Xiao Hu}
\affiliation{%
  \department{Center for Data Science and Nell Hodgson Woodruff School of Nursing}
  \institution{Emory University}
  \city{Atlanta}
  \state{Georgia}
  \country{USA}
}
\email{xiao.hu@emory.edu}

\author{Runze Yan}
\affiliation{%
  \department{Center for Data Science and Nell Hodgson Woodruff School of Nursing}
  \institution{Emory University}
  \city{Atlanta}
  \state{Georgia}
  \country{USA}
}
\email{runze.yan@emory.edu}

\renewcommand{\shortauthors}{Liu et al.}

\begin{abstract}
Image-based dietary assessment can automate initial food logging, but meal photographs often leave food identity, hidden ingredients, preparation, and portion uncertain, making correction a recurring part of use. We introduce \emph{Human--AI Collaborative Correction}, which reframes correction as \emph{communicate intent $\rightarrow$ regenerate $\rightarrow$ verify}: users express what should change through text, voice, or guided clarification, the AI regenerates the linked dietary record, and users verify the result. In a within-participant study with 30 participants, compared with conventional \emph{Manual Search-and-Edit Correction}, collaborative correction was associated with lower workload (Raw TLX: 35.83 vs.\ 55.00; $p<.001$), 21.0\% lower recorded usage time, 51.1\% fewer recorded clicks, and more favorable
user-experience ratings; 90.0\% preferred it overall. An exploratory nutrient analysis also found lower energy MAE. These findings support intent-driven regeneration as a promising interaction pattern for reducing correction burden while preserving user verification and making image-based dietary tracking easier to sustain.
\end{abstract}



\begin{teaserfigure}
  \centering
  \includegraphics[width=0.9\textwidth]{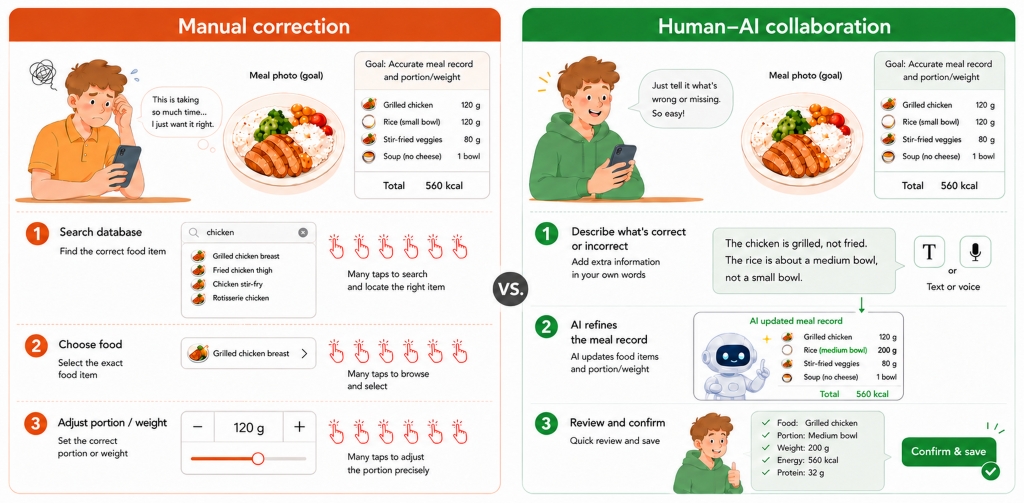}
  \caption{Two interaction modes for correcting an AI-generated dietary record. Manual Search-and-Edit Correction requires users to translate what they know about the meal into database searches, entry selection, and field-level portion or weight edits. Human--AI Collaborative Correction reframes this workflow as \emph{communicate $\rightarrow$ regenerate $\rightarrow$ verify}: users express the intended semantic correction, the AI regenerates the linked record, and users verify the revision.}
  \Description{Side-by-side illustration of two meal-correction workflows. On the left, a user manually searches a food database, selects a food entry, and adjusts its portion or weight. On the right, a user communicates by text or speech that the chicken is grilled and the rice portion is a medium bowl; the AI regenerates the meal record, and the user verifies it.}
  \label{fig:correction-concept}
\end{teaserfigure}

\maketitle

\section{Introduction}

Food logging has traditionally required people to construct dietary records item by item: recalling what they ate, locating suitable database entries, and specifying amounts \cite{cordeiro2015barriers,luo2019codesigning}. Because food logging requires repeated effort in daily life, high recording burden can make sustained tracking difficult \cite{cordeiro2015barriers,epstein2015lived}. Image-based dietary assessment changes where this work begins. Earlier systems supported photographic capture and user review or delegated nutritional analysis to crowd workers \cite{kim2010mobileui,noronha2011platemate}; more recent AI-based systems can automatically generate an initial dietary record of food identities, portions, and nutrients from a meal photograph \cite{phalle2025navigating, shonkoff2023ai, chotwanvirat2024advancements}. This AI-driven automation reduces the work of constructing an initial record, but it does not eliminate the need for user involvement. The practical usefulness of an AI-generated record still depends on whether users can recognize and correct what the image-based estimate gets wrong or leaves unresolved.

Meal photographs provide only partial evidence about what was eaten. Foods with different compositions can look alike, ingredients may be obscured within mixed dishes, and a two-dimensional image provides limited information about portion size \cite{fridolfsson2025performance,yan2025dietai24}. Preparation details may also depend on context unavailable from appearance alone, such as cooking state or the use of added oil \cite{chen2025diabot}. Customized multimodal models can improve food recognition \cite{chen2025diabot}, but some uncertainties arise not simply from model error, but because relevant information is ambiguous or absent from the image itself \cite{chen2025diabot,fridolfsson2025performance}. The person who ate the meal may know which milk was used, whether oil was added, what a filling contains, or approximately how much food was served. Because such uncertainties can recur across logging occasions, correction can become a recurring part of image-based dietary assessment rather than a one-off recovery step: the system produces a provisional record, and the user may need to supply information that the image cannot provide.

A common database-based approach to correction is what we call \emph{Manual Search-and-Edit Correction}, or \emph{manual correction}: users search for and select food entries, add or remove foods, and edit portions or weights \cite{cordeiro2015barriers,evenepoel2020myfitnesspal}. Even when users already know what needs to change, they must translate that meal knowledge into the operations required by a structured food database. Prior food-journaling research documents ambiguous database entries and difficulties representing meals at the intended level of detail \cite{cordeiro2015barriers,luo2019codesigning}, while conversational food-journaling research suggests opportunities for people to express meal information in their own terms \cite{silva2025foodytalk}. The resulting gap is one of translation: knowing that vegetables were cooked with oil is different from locating an appropriate database entry, choosing a representation and unit, and editing the corresponding fields. When corrections recur across meals, this translation work compounds the repeated burden of dietary tracking. We interpret this mismatch through Hutchins et al.'s gulf of execution \cite{hutchins1985direct}: possessing the relevant meal knowledge does not necessarily make it easy to act on a structured dietary record.

Motivated by this translation burden, we ask whether correction can instead be organized around communicating what should change rather than manually carrying out each database operation. We introduce \emph{Human--AI Collaborative Correction} (\emph{collaborative correction}), an interaction model organized around \emph{communicate intent $\rightarrow$ regenerate $\rightarrow$ verify}. Users communicate the intended correction, the AI translates that intent into updates to the linked dietary record, and users inspect the regenerated result before accepting it. Rather than removing users from correction, this model reallocates correction work: people provide meal knowledge and retain control over regeneration and acceptance, while the AI performs the translation into linked food, portion, and nutrient fields.

\begin{samepage}
We examine three parallel outcome domains:

\begin{description}
  \item[RQ1 --- Correction Burden:] How does perceived correction burden differ between Human–AI Collaborative Correction and Manual Search-and-Edit Correction?
  \item[RQ2 --- Correction Efficiency and User Experience:] How do the two correction methods differ in correction efficiency and user experience?
  \item[RQ3 --- Post-Correction Nutrient Accuracy:] How do the two correction methods differ in energy and macronutrient error relative to the reference values?
\end{description}
\end{samepage}

Figure~\ref{fig:correction-concept} contrasts the two interaction modes. Manual correction requires users to translate an intended change into database search, entry selection, and field-level editing, whereas collaborative correction follows \emph{communicate intent $\rightarrow$ regenerate $\rightarrow$ verify}. This allocation draws on mixed-initiative interaction and guidance on efficient correction and user control \cite{horvitz1999mixedinitiative,amershi2019guidelines}. Free-form language alone, however, can be difficult to formulate and its effects difficult to anticipate \cite{subramonyam2024gulf,zamfirescu2023johnny}; structured refinement controls and inspectable outputs can provide complementary support \cite{wu2022aichains,wang2024promptcharm}. Verification remains substantive user work because people must determine whether regeneration fulfilled their intent \cite{tankelevitch2024metacognitive}.

To instantiate and evaluate this interaction model, we developed a research prototype with three complementary mechanisms for communicating correction intent. \emph{Text/Voice Feedback} lets users describe intended corrections in their own words; \emph{Quick Clarification} presents audit-generated candidate corrections for selection; and \emph{Visualized Portion Size} provides visual references for communicating an intended amount. These mechanisms may be used individually or combined within the same correction episode; they are not separate experimental conditions. Selected inputs are assembled as correction context, while users explicitly decide when to initiate regeneration and whether to accept the returned record. The prototype therefore enables us to examine how delegating linked record updates to AI changes the work of expressing, implementing, and verifying dietary corrections.

We conducted a within-participant study with 30 participants using predefined record errors across five recurring image-based dietary-assessment challenges. Manual correction preceded collaborative correction for all participants, while the two non-overlapping image sets were rotated between methods across participants. Participants were provided food identity, composition, and preparation information and judged portions from the photographs, allowing the study to focus on implementing known corrections rather than independently detecting every error. In the paired survey responses, collaborative correction received lower perceived workload ratings and more favorable ratings of ease, confidence, perceived control, trust, and intended everyday use. An exploratory behavioral analysis showed 21.0\% lower mean total recorded usage time (251.05 vs.\ 317.77~s) and 51.1\% fewer mean recorded clicks (21.36 vs.\ 43.68; both $p_{\mathrm{Holm}}<.001$; Figure~\ref{fig:combined-results}(a)). The nutrient-level analysis found lower energy error with collaborative correction across the ten study photographs.

This paper makes three contributions:
\begin{enumerate}
  \item \textbf{An interaction model.} We introduce Human--AI Collaborative Correction, which organizes correction as \emph{communicate intent $\rightarrow$ regenerate $\rightarrow$ verify}. The model shifts the user's role from manually translating meal knowledge into database operations toward communicating what the record should represent, while retaining explicit control over regeneration and acceptance.

  \item \textbf{A human-centered research prototype.} We instantiate this interaction model through free-form Text/Voice Feedback, selectable Quick Clarification candidates, and Visualized Portion Size cues, providing complementary ways to communicate correction intent while preserving explicit user authorization and verification.

  \item \textbf{A controlled empirical comparison.} We compare Human--AI Collaborative Correction with Manual Search-and-Edit Correction using fixed method order and image sets rotated between methods across five recurring image-based dietary-assessment challenges, examining correction burden, behavioral efficiency, user experience, and post-correction nutrient accuracy.
\end{enumerate}

\section{Related Work}

\subsection{From Manual Logging to Semi-Automated Dietary Assessment}

Food logging places repeated demands on the collection and interpretation of personal data. Li et al.'s stage-based model of personal informatics describes how barriers in collection and integration can carry forward into later reflection and action \cite{li2010stagebased}, while lived informatics situates tracking within changing routines, motivations, and lapses \cite{epstein2015lived}. In food journaling, these demands include locating suitable database entries and translating meals into standardized foods, units, and amounts \cite{cordeiro2015barriers}. Lightweight photo-based capture can reduce the need for exhaustive quantitative entry while preserving opportunities for reflection \cite{cordeiro2015rethinking}. More broadly, semi-automated tracking seeks to balance the roles of people and automation: automated sensing can reduce capture work, while users provide contextual information that is difficult for a system to infer \cite{choe2017semiautomated}. This perspective shifts attention from how much dietary logging can be automated to how the remaining user work should be supported.

Even when dietary data collection is partially automated, users still need ways to provide meal details that the system may not capture and to ensure that those details are represented appropriately in the dietary record. Co-design with dietitians has identified needs for flexibility in food-record content and granularity, while Foody Talk explores conversational journaling that allows people to describe meals in their own terms \cite{luo2019codesigning,silva2025foodytalk}. Together, this work suggests that reducing logging burden involves not only shortening data entry, but also making it easier to translate people's knowledge of a meal into a structured dietary record. Earlier image-based systems explored different ways to reduce this work: PlateMate used crowdsourcing to estimate nutritional information from food photographs, whereas Kim et al. developed an interface that returned automated food-recognition and volume estimates for user review \cite{noronha2011platemate,kim2010mobileui}. Benchmark datasets subsequently enabled more systematic development and evaluation of automated dietary assessment. Nutrition5k links visual data with component weights and nutritional annotations, while SNAPMe links smartphone meal photographs from everyday eating contexts with independently collected food records from the Automated Self-Administered 24-Hour (ASA24) Dietary Assessment Tool \cite{thames2021nutrition5k,larke2023snapme}. Complementing these technical advances, Chung et al.'s observation and interview study of dietitians reviewing photo diaries highlights the importance of contextual information, eating patterns, variety, and portion when interpreting dietary images, and points toward opportunities to combine visual analysis with contextual information and expert--client collaboration \cite{chung2024diaries}.

Recent multimodal AI work further extends this trajectory through large-scale training resources, evaluation of general-purpose models, customized dietary assistants, and database-grounded assessment. NutriMLLM uses large-scale synthetic image--description--nutrient data to fine-tune multimodal large language models (MLLMs) for comprehensive nutrient estimation \cite{yan2026nutrimllm}. Fridolfsson et al. report portion-dependent underestimation and variability in nutrient estimates across three multimodal models, while Chen et al.'s Diabot-GPT-4o study examines customization and food-name input and identifies continuing challenges involving portion, visibility, and prompting \cite{fridolfsson2025performance,chen2025diabot}. DietAI24 provides a complementary approach by grounding multimodal dietary assessment in standardized food and portion codes \cite{yan2025dietai24}. These advances improve the generation of an initial dietary record, but estimation accuracy and overall system usability do not characterize the work required when that record must be corrected. Prior image-based systems have supported user review and, in some cases, correction---for example, search-based replacement of misidentified food labels in Kim et al.'s interface \cite{kim2010mobileui}. However, existing work has primarily focused on capture, recognition, estimation, benchmarking, and overall system performance rather than on correction as an interaction problem. It therefore remains unclear how different correction workflows affect the effort required to translate users' meal knowledge into changes to an AI-generated dietary record. Our work addresses this gap by comparing manual field-level correction with intent-driven correction under controlled starting errors and meal information.

\subsection{Correcting AI Outputs: Direct Manipulation, Delegation, and Generative Steering}

Correction can allocate work between people and automated systems in different ways. Hutchins et al.'s gulfs of execution and evaluation distinguish the effort required to translate an intention into system actions from the effort required to determine whether those actions produced the intended result \cite{hutchins1985direct}. Shneiderman and Maes describe a related tension between direct manipulation, which emphasizes visible and predictable user actions, and interface agents that act on a user's behalf \cite{shneiderman1997direct}. Mixed-initiative principles and human--AI interaction guidelines further consider how action costs, uncertainty, efficient correction, and user control should shape this allocation \cite{horvitz1999mixedinitiative,amershi2019guidelines}.

Correction itself has long been studied in human--computer interaction (HCI). Multimodal speech interfaces provide alternative modalities for repairing recognition errors, while machine-translation post-editing research examines the time and quality associated with repairing generated outputs \cite{suhm2001multimodal,green2013postediting}. More recently, DirectGPT combines direct manipulation with language-model revision, allowing users to specify targets for localized generative changes \cite{masson2024directgpt}. These systems demonstrate that correction is not simply an afterthought to generation: how users specify a change, how the system carries it out, and how the result is returned for inspection are interaction-design choices in their own right.

Generative systems introduce an additional form of correction through natural-language steering. This can reduce the need to specify low-level operations, but it introduces the work of formulating instructions and anticipating their effects. Prompting research documents difficulties in expressing intent, while the gulf of envisioning describes challenges in anticipating what a generative system will produce from an instruction \cite{zamfirescu2023johnny,subramonyam2024gulf}. Interfaces such as AI Chains expose intermediate operations and outputs, and PromptCharm combines multimodal prompting with structured refinement controls \cite{wu2022aichains,wang2024promptcharm}. Together, this work suggests that intent-driven correction should not rely on free-form prompting alone: selectable alternatives, structured refinement, and inspectable outputs can help users communicate and evaluate intended changes. Our work applies this perspective to dietary records, where a correction expressed at the semantic level may require coordinated updates to multiple structured fields.

\subsection{Verification, Control, and Appropriate Reliance}

Delegating correction to AI shifts some work from executing changes to evaluating their consequences. This distinction is particularly important because subjective judgments of an AI system and objective correctness can diverge. Trust supports appropriate reliance only in relation to a system's actual capabilities and context \cite{leesee2004trust}. Kocielnik et al. showed that systems with the same overall accuracy but different error profiles can produce different perceptions of accuracy and acceptance, and that expectation-setting can influence acceptance \cite{kocielnik2019imperfect}. Explanations can increase acceptance of incorrect as well as correct recommendations, while interventions that reduce overreliance may receive less favorable subjective evaluations \cite{bansal2021whole,bucinca2021trust}. Confidence, trust, preference, and perceived ease therefore cannot substitute for independent assessment of the resulting record.

Human--AI interaction guidelines emphasize support for correction and disambiguation, while work on the metacognitive demands of generative AI identifies monitoring and evaluation as substantive user work \cite{amershi2019guidelines,tankelevitch2024metacognitive}. Verification should consequently be treated as an interaction stage rather than inferred from the mere presence of a confirmation action. Related interfaces such as DynaVis and Data Formulator connect high-level user intent to structured changes through editable controls and inspectable representations \cite{vaithilingam2024dynavis,wang2024dataformulator}. These systems illustrate ways of combining generative transformation with opportunities to examine and refine its effects.

Dietary records make this verification problem especially concrete because their fields are interdependent: food identity determines available portion options, portions map to gram weights, and food identity and amount jointly determine nutrient estimates \cite{yan2025dietai24}. A semantic correction to one aspect of a meal can therefore propagate to other fields that the user did not directly edit. This creates a tradeoff: delegating linked updates may reduce the work of carrying out a correction, but it also makes inspection of the regenerated record consequential. Prior work establishes important principles for automation, correction, generative steering, and appropriate reliance, but does not establish how manual field-level correction compares with intent-driven regeneration for structured dietary records. We address this gap by comparing the two interaction modes under controlled starting errors and meal information, while separately evaluating subjective experience, behavioral efficiency, and post-correction nutrient accuracy.

\section{Human--AI Collaborative Correction: Design and Prototype}
\label{sec:system}

\subsection{Interaction Model}

Human--AI Collaborative Correction organizes dietary-record repair as \emph{communicate intent $\rightarrow$ regenerate $\rightarrow$ verify}. Users communicate what should change in the current record; the AI translates that intent into coordinated updates to linked food, portion, and nutrient fields; and users inspect the regenerated record before accepting it or requesting another revision. We developed a research prototype to instantiate and evaluate this allocation of correction work.

\subsection{Design Rationale}

The design shifts users from specifying database operations toward communicating meal knowledge. This addresses the gulf of execution by allowing people to state what the record should represent rather than requiring them to determine how to implement the change through database search, entry selection, and field-level editing \cite{hutchins1985direct}. The resulting division of work follows principles of semi-automated and mixed-initiative interaction: people contribute contextual knowledge that may be unavailable to the system, while automation carries out the corresponding structured record updates \cite{choe2017semiautomated,horvitz1999mixedinitiative}.

Communicating intent, however, should not depend on free-form prompting alone. Formulating an instruction and anticipating its effects can be difficult \cite{zamfirescu2023johnny,subramonyam2024gulf}. We therefore combine free-form input with selectable correction candidates and visual portion references.
These inputs assemble correction context without immediately modifying the dietary record; only explicit submission initiates regeneration. The regenerated record is then returned for inspection before acceptance. This separates authorization to regenerate from authorization to accept and applies human--AI
guidance on efficient correction, disambiguation, and user control \cite{amershi2019guidelines}.

\subsection{Expressing Correction Intent}

As summarized in Figure~\ref{fig:architecture}, the prototype provides three complementary mechanisms for communicating correction intent: \textbf{Text/Voice Feedback}, \textbf{Quick Clarification}, and
\textbf{Visualized Portion Size}. Users may use these mechanisms individually or combine them within a correction request; they are not separate experimental conditions. \textbf{Text/Voice Feedback} supports free-form expression: users can describe what is wrong or what should change in their own words through typed text or speech. This is particularly useful when relevant information is difficult to infer from the meal image or capture through predefined options, such as hidden ingredients, preparation details, or other contextual information known to the eater.

\textbf{Quick Clarification} supports recognition rather than recall: users may find it easier to select an appropriate correction from plausible alternatives than to formulate a complete correction from scratch. This mechanism builds on a broader use of LLMs as critics that inspect an initial model output and generate feedback or candidate revisions for subsequent refinement \cite{madaan2023self,gou2024critic,ke2024critiquellm}. We adapt this critique paradigm to recurring uncertainties in image-based dietary assessment. Rather than allowing model-generated critique to autonomously revise the record, Quick Clarification surfaces plausible alternatives for the user to consider. The user decides whether any candidate should become part of the correction request, placing human judgment between model critique and regeneration.

\textbf{Visualized Portion Size} supports communication of an intended amount through visual references rather than requiring users to specify a numerical weight. Users compare alternative portion representations and select the one that best reflects their intended correction. This selection can be combined with Text/Voice Feedback or Quick Clarification when no single mechanism fully expresses the intended change. Together, the three mechanisms support free-form
expression, recognition of plausible alternatives, and visual specification of portion size. Their concrete implementation and integration with the dietary-assessment engine are described next.

\subsection{Prototype Architecture and Implementation}
\label{sec:prompt-workflow}
\label{sec:prototype-implementation}

\begin{figure}[t]
  \centering
  \includegraphics[width=\linewidth]{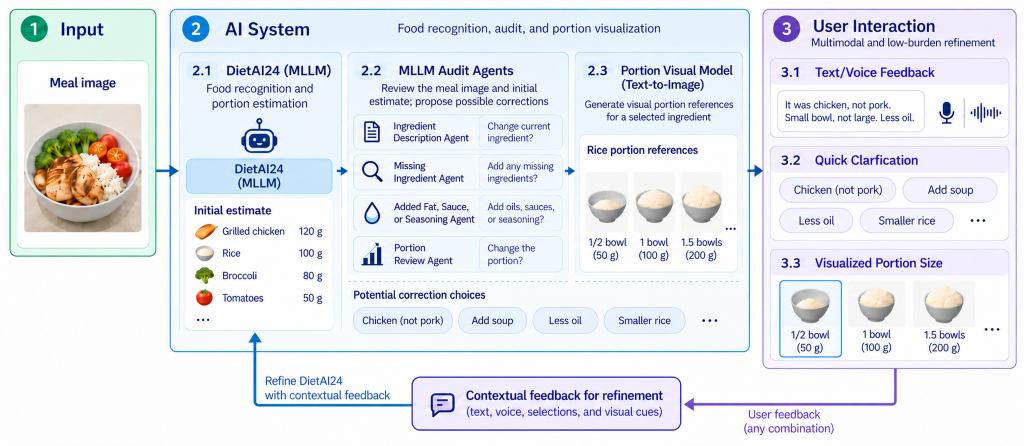}
  \caption{Architecture of Human--AI Collaborative Correction. The existing
  DietAI24 dietary-assessment pipeline generates the provisional dietary record.
  Text/Voice Feedback, audit-generated Quick Clarification candidates, and
  Visualized Portion Size references provide complementary correction inputs.
  User-selected inputs are assembled as additional context and passed back to
  DietAI24 for a subsequent assessment pass rather than directly modifying
  record fields. The regenerated record is returned for user verification and
  acceptance or further correction. Food quantities in the artwork are
  illustrative.}
  \Description{The artwork shows a meal image entering DietAI24 for an
  initial dietary assessment, four audit agents proposing correction candidates,
  a portion-visualization module, and user controls for text or voice, candidate
  selection, and visual portion selection. Selected correction information is
  assembled as additional context and returned to DietAI24.}
  \label{fig:architecture}
\end{figure}

Figure~\ref{fig:architecture} shows how Human--AI Collaborative Correction is implemented around \textbf{DietAI24}, an existing public MLLM-based framework for image-based dietary assessment \cite{yan2025dietai24}. We selected DietAI24 because it already provides an end-to-end multimodal dietary-assessment pipeline that translates meal inputs into standardized food, portion, and nutrient estimates. This allowed us to study correction as an interaction problem without developing or modifying the underlying dietary-assessment model. We therefore reuse DietAI24's existing estimation and database-grounding pipeline as the assessment backend. Human--AI Collaborative Correction operates as an additional interaction layer: correction information supplied or selected by the user is assembled as additional context and passed back to DietAI24, which performs another assessment pass to refine the dietary record.

DietAI24's existing pipeline uses retrieval-augmented generation (RAG) to ground MLLM outputs in the USDA Food and Nutrient Database for Dietary Studies (FNDDS). The reused DietAI24 backend in our study used GPT-5.4 within this existing RAG pipeline. DietAI24 uses FNDDS 2019--2020, which contains 5,624 food items and more than 23,000 associated portion options \cite{yan2025dietai24,rhodes2022fndds}. Foods are represented by FNDDS food codes together with associated portion descriptions and gram weights, and food identity and amount jointly determine the corresponding nutrient estimates \cite{yan2025dietai24}. This structured representation provides the basis for DietAI24 to update linked food, portion, weight, and nutrient information when additional correction context is supplied.

The three correction mechanisms contribute complementary forms of context to this refinement process. Text/Voice Feedback contributes free-form text or voice transcription. Quick Clarification is implemented using four GPT-5.4 MLLM audit agents, each focusing on a recurring source of uncertainty in image-based dietary assessment. The \emph{Ingredient Confirmation Agent} proposes alternative descriptions for a specified food; the \emph{Missing Ingredient
Agent} proposes potentially omitted foods; the \emph{Added Fat, Sauce, or Seasoning Agent} proposes plausible fats, sauces, dressings, condiments, or seasonings; and the \emph{Portion Review Agent} proposes alternative amounts. Each agent ranks plausible candidates for presentation to the user. The audit agents do not directly modify the dietary record; only candidates selected by the user are incorporated into the correction context. Visualized Portion Size is implemented using GPT-Image-2 and the standardized portion information available through the existing DietAI24/FNDDS pipeline. GPT-Image-2 receives an FNDDS food-code description and an associated FNDDS portion-size description and generates a visual reference for that standardized portion. A selected visual reference contributes its associated portion information to the correction context. The complete prompts for the four audit agents and visual-cue generator appear in Appendix~\ref{app:prompts}.

The client combines free-form text or voice transcription, user-selected Quick Clarification candidates, selected portion information, or a combination of these inputs into a common correction context. Selecting or assembling these inputs does not itself alter the current dietary record. Only explicit user submission sends the correction context, together with the meal input and current record, back to DietAI24. DietAI24 then performs another dietary-assessment pass using the additional user-provided context and returns a refined structured record with updated food descriptions and codes, portion representations, gram weights, and nutrient estimates. The regenerated record is presented for inspection, with ``View original'' supporting comparison with the previous record; users may accept the result or provide further correction intent and repeat the loop. We implemented this collaborative correction workflow in NutriCamp (\url{https://nutricamp.ai/}), a mobile dietary-assessment application developed by our team. The NutriCamp implementation integrates the three correction mechanisms with the existing DietAI24 assessment pipeline, without modifying the underlying dietary-assessment model. NutriCamp served as the research prototype evaluated in this study, and Figure 3 presents screenshots of its correction interface.

\subsection{Verification and End-to-End Walkthrough}

Verification is consequential because regeneration can affect more than the field explicitly targeted by the user. Changing a food identity, for example, can change the applicable portion descriptions, gram weight, and derived nutrient estimates; regeneration may also alter information that the user did
not intend to change. The prototype therefore presents the regenerated record for inspection before acceptance, allowing users to check both the requested correction and linked fields that should remain consistent. Users may accept the record or communicate additional intent and repeat the correction loop.
Participant confirmation ends the interaction; acceptance alone does not establish the accuracy of the saved record.

\begin{figure*}[t]
  \centering
  \begin{minipage}[t]{0.27\textwidth}
    \centering
    \includegraphics[trim=146bp 120bp 146bp 150bp,clip,width=0.792\linewidth]{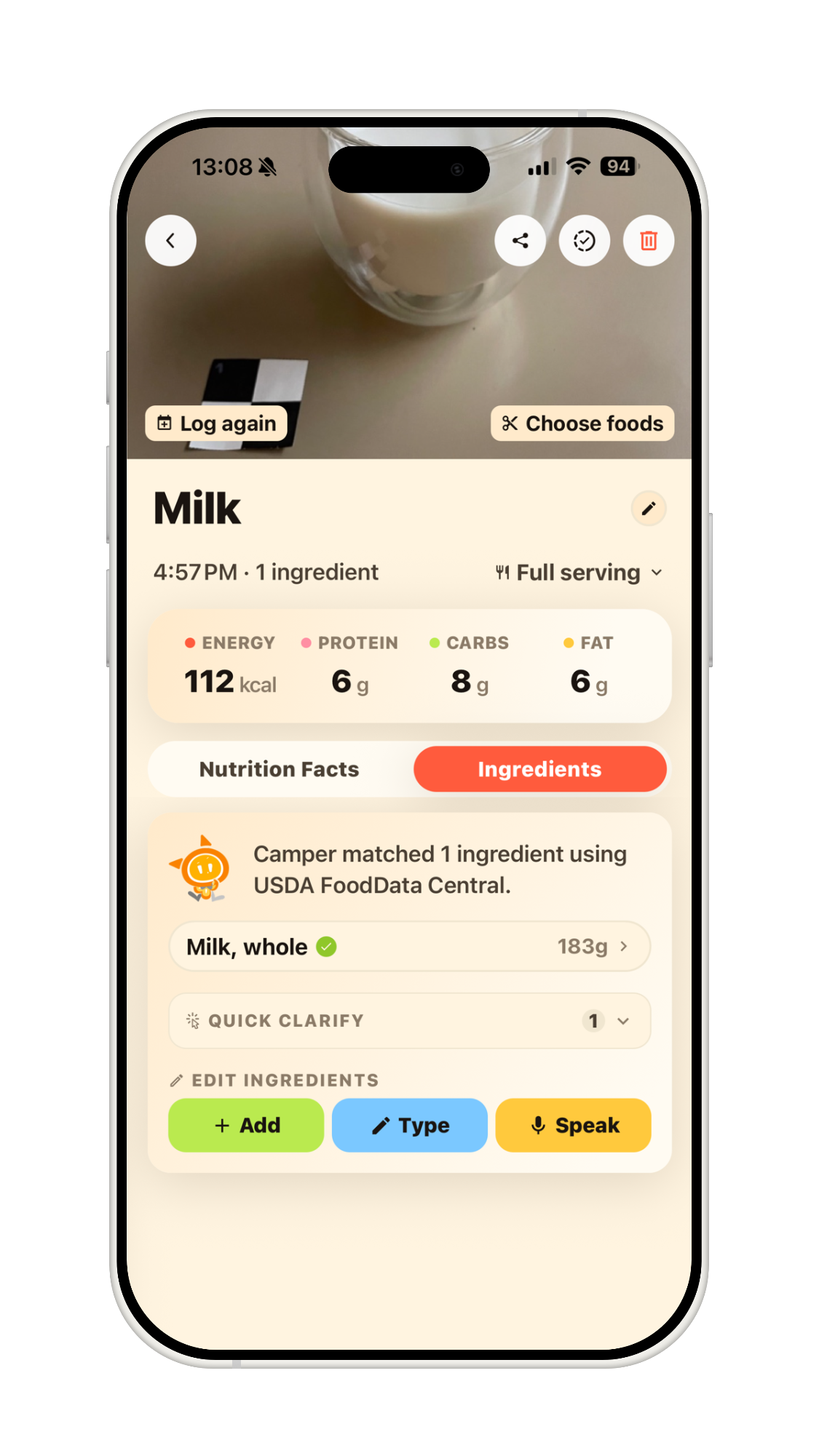}
    \par\smallskip
    {\footnotesize\textbf{(A) Initial result}\par}
  \end{minipage}\hspace{0.025\textwidth}
  \begin{minipage}[t]{0.27\textwidth}
    \centering
    \includegraphics[trim=146bp 120bp 146bp 150bp,clip,width=0.792\linewidth]{\detokenize{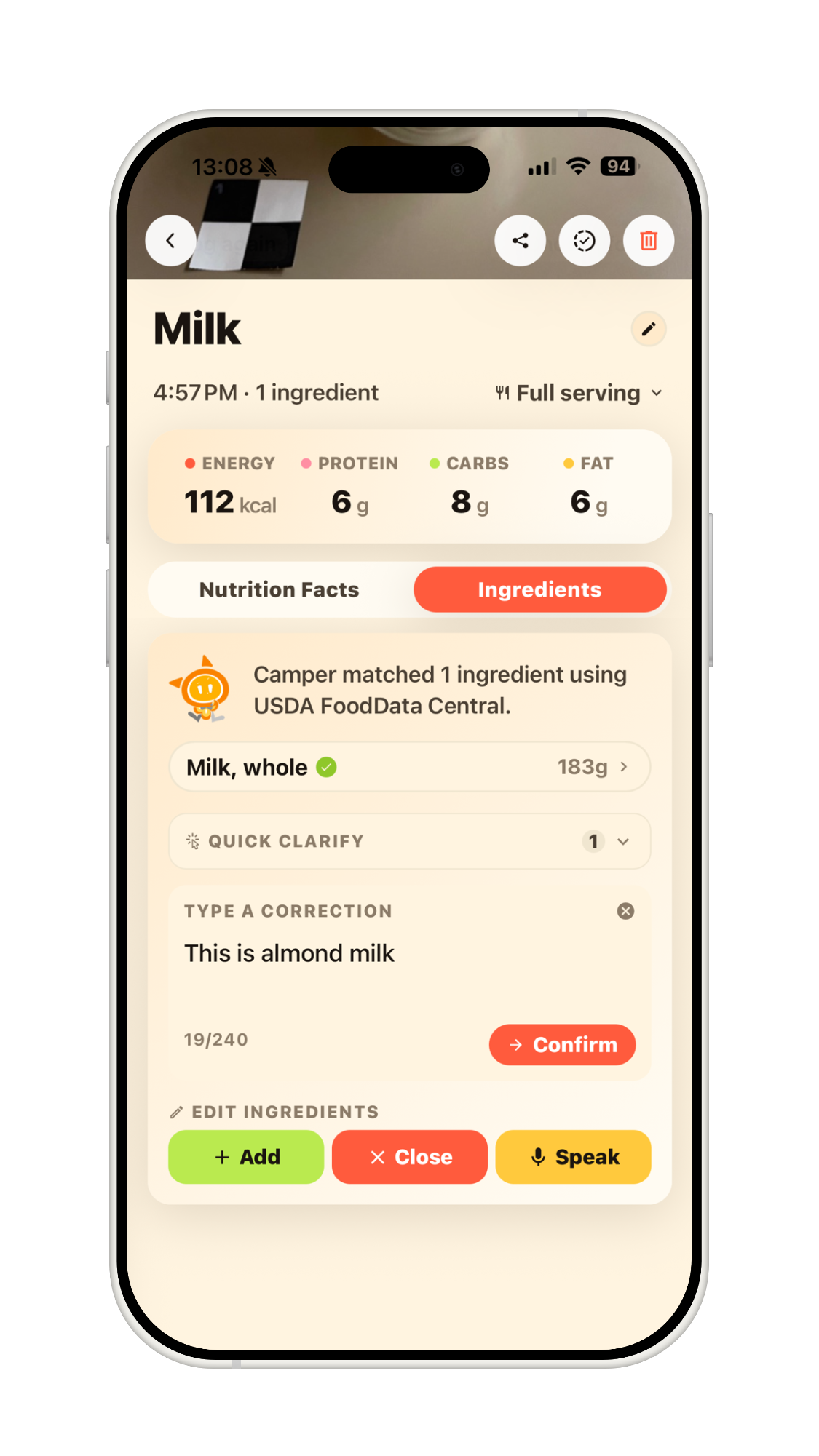}}
    \par\smallskip
    {\footnotesize\textbf{(B) Text/Voice\\Feedback}\par}
  \end{minipage}\hspace{0.025\textwidth}
  \begin{minipage}[t]{0.27\textwidth}
    \centering
    \includegraphics[trim=146bp 120bp 146bp 150bp,clip,width=0.792\linewidth]{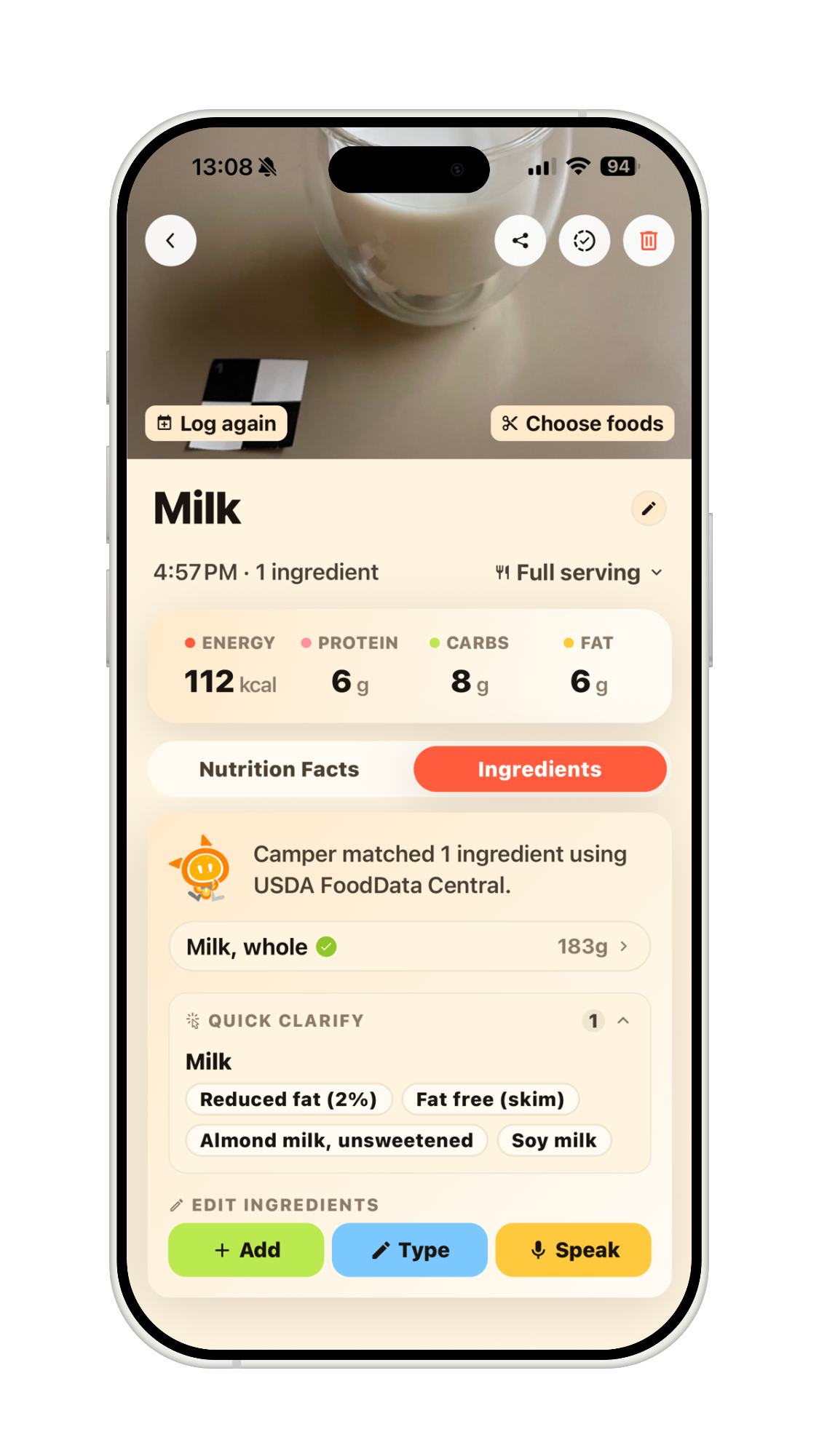}
    \par\smallskip
    {\footnotesize\textbf{(C) Quick\\Clarification}\par}
  \end{minipage}
  \par\medskip
  \begin{minipage}[t]{0.27\textwidth}
    \centering
    \includegraphics[trim=146bp 120bp 146bp 150bp,clip,width=0.792\linewidth]{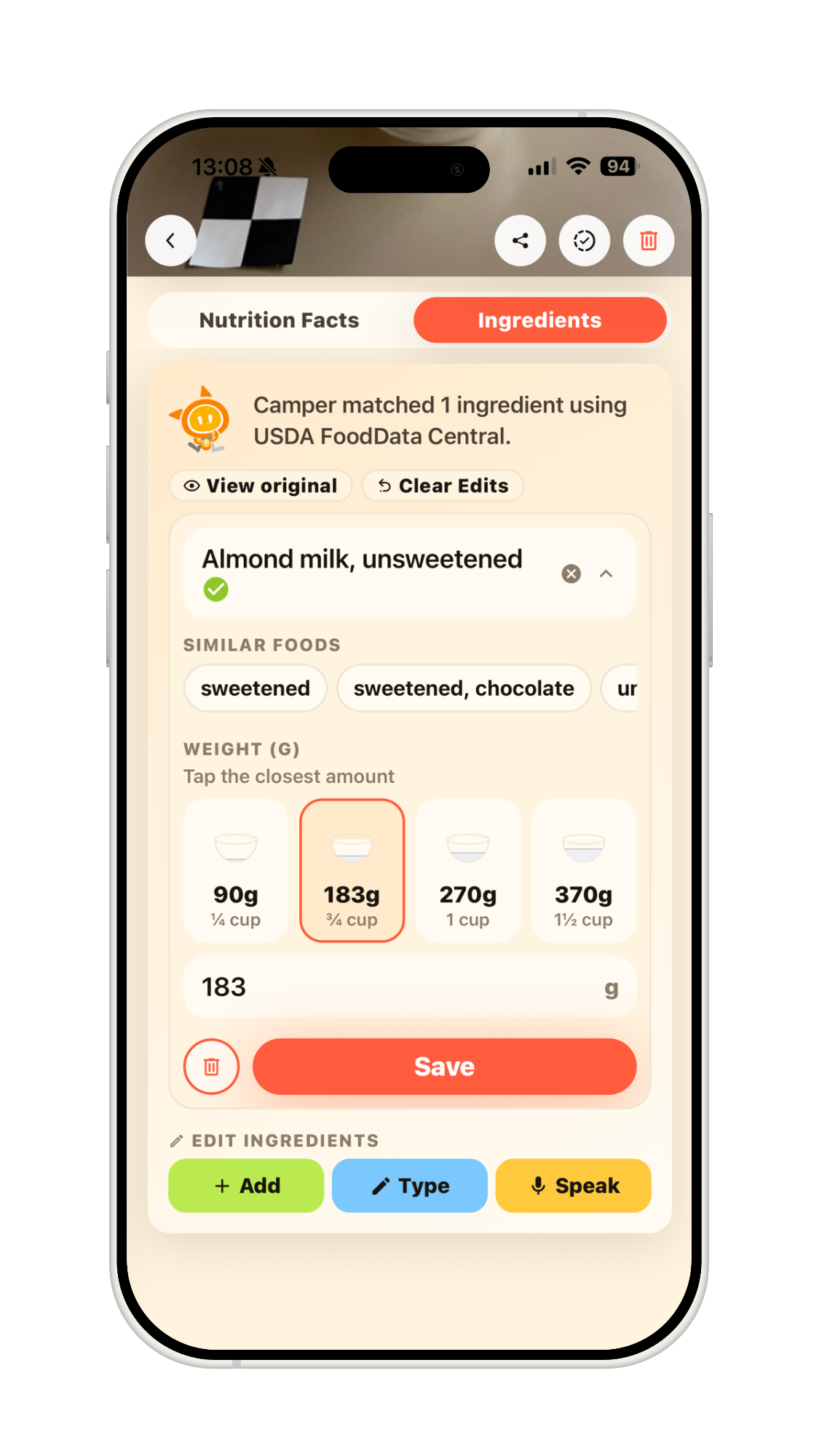}
    \par\smallskip
    {\footnotesize\textbf{(D) Visualized\\Portion Size}\par}
  \end{minipage}\hspace{0.025\textwidth}
  \begin{minipage}[t]{0.27\textwidth}
    \centering
    \includegraphics[trim=146bp 120bp 146bp 150bp,clip,width=0.792\linewidth]{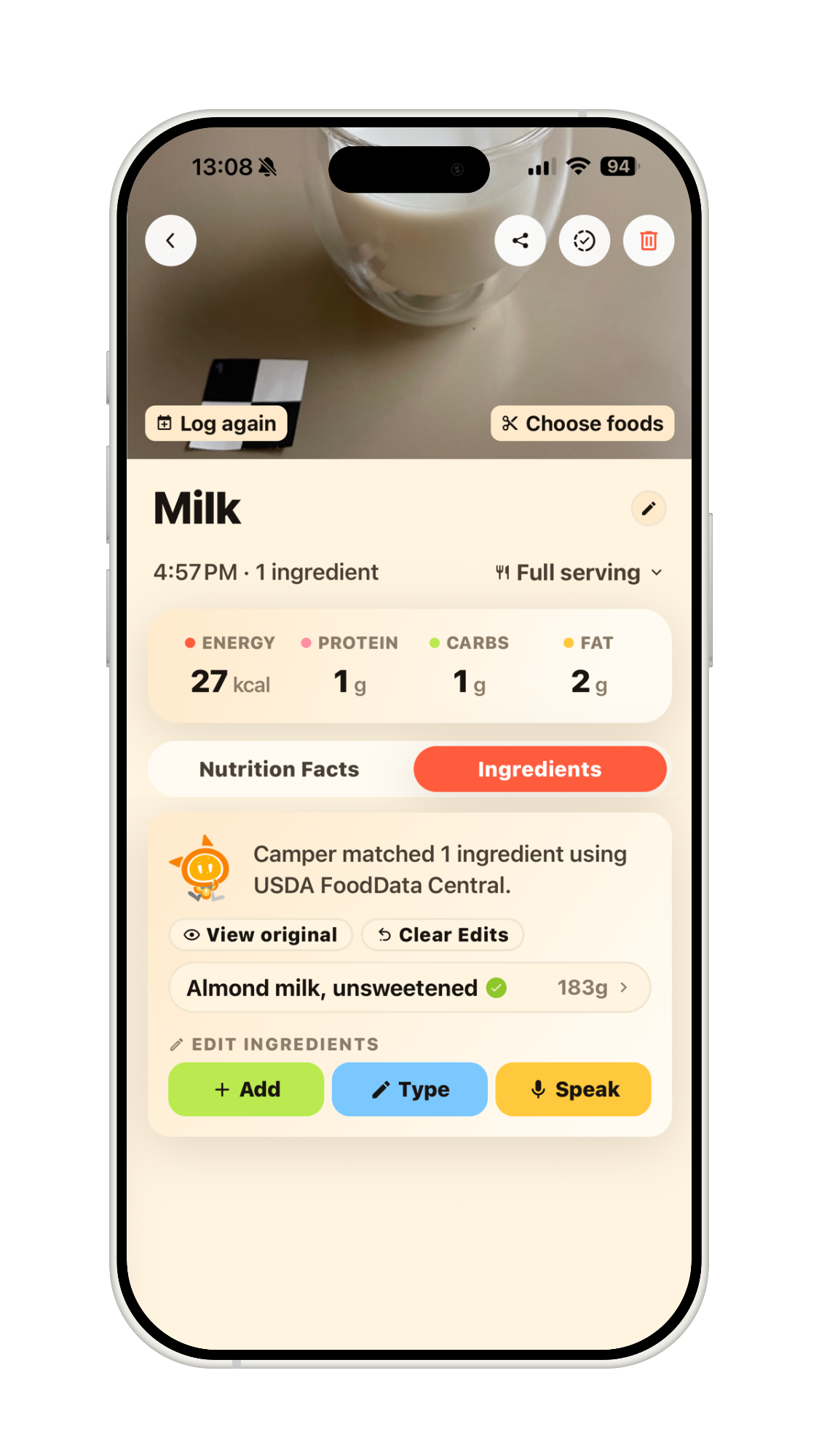}
    \par\smallskip
    {\footnotesize\textbf{(E) Corrected record}\par}
  \end{minipage}
  \caption{Screenshots from NutriCamp, our developed mobile dietary-assessment application, illustrating the Human–AI Collaborative Correction workflow for correcting whole milk to unsweetened almond milk: (A) initial dietary record; (B) typed correction through Text/Voice Feedback; (C) Quick Clarification candidates; (D) Visualized Portion Size references, with 183\,g highlighted; and (E) regenerated record for verification. Panels B--D illustrate complementary correction resources rather than separate experimental conditions. During onboarding, participants were encouraged to check Quick Clarification first, use text or speech when an appropriate correction was not available, and adjust ingredient or portion options as needed; this was a suggested workflow rather than a mandatory sequence. The regenerated record retains the 183\,g weight while updating food identity and nutrient estimates.}
  \Description{Five smartphone screenshots of a milk-record correction. A shows whole milk at 183 grams and 112 kilocalories. B shows the typed correction ``This is almond milk'' and a Confirm button. C shows Quick Clarify candidates including reduced-fat milk, skim milk, unsweetened almond milk, and soy milk. D shows the unsweetened almond milk entry with visual portion options of 90, 183, 270, and 370 grams; 183 grams is highlighted above a Save button. E shows the regenerated unsweetened almond milk record at 183 grams and 27 kilocalories, with View original and Clear Edits controls.}
  \label{fig:prototype-screens}
\end{figure*}

Figure~\ref{fig:prototype-screens} provides an illustrative end-to-end walkthrough of the collaborative correction workflow, separate from the study-task reference values. In this example, a beverage initially recorded as 183 g of whole milk is changed to unsweetened almond milk. The user can communicate the correction through Text/Voice Feedback (B) or select unsweetened almond milk from the Quick Clarification
candidates (C). The Visualized Portion Size interface provides visual references for the corrected food, with 183\,g highlighted (D). The regenerated record (E) shows unsweetened almond milk at 183\,g with updated energy and macronutrient estimates. The ``View original'' control allows comparison with the initial record before the revision is accepted.

\section{Study Design}
\label{sec:study-design}

\subsection{Study Design and Experimental Controls}
\label{sec:starting-records}

We used a within-participant design to compare
\emph{Human--AI Collaborative Correction} with
\emph{Manual Search-and-Edit Correction}, hereafter
\emph{collaborative correction} and \emph{manual correction}. In collaboration
with dietitians, we identified five recurring challenges in image-based dietary
assessment that frequently require user correction or verification:
\emph{visual similarity}, where foods with different identities may look alike;
\emph{hidden composition}, where ingredients are obscured inside or beneath
other foods; \emph{mixed dishes}, where multiple components must be identified
and represented; \emph{preparation}, where attributes such as added oil or
cooking method may not be visually apparent; and \emph{portion}, where the
amount of an otherwise identified food may be uncertain.

Before each formal trial, participants viewed the meal photograph and a food
card specifying the food identity, composition, and task-relevant preparation;
the reference portion was withheld. Participants therefore did not need to
infer food identity or preparation, but still judged portion from the
photograph. This design focused the study on implementing known corrections
rather than independently detecting errors. Each participant completed ten
distinct tasks: five per method and one from each challenge in each condition.
The images formed two five-image sets, A1--A5 and B1--B5, each containing one
example of every challenge. The sets were rotated between methods across
participants so that no participant corrected the same photograph with both
methods. For each challenge, the two conditions used matched predefined errors
and the same correction objective to support a fair comparison. Tasks followed
the fixed challenge order shown in Figure~\ref{fig:stimuli}. Table~\ref{tab:stimuli} summarizes the standardized starting errors and correction targets.

Method order was fixed, with manual correction preceding collaborative correction for all participants. To limit stimulus-specific carryover, no participant corrected the same photograph with both methods. Instead, the two image sets were rotated between methods across participants, so that each set appeared with both correction methods across the study and neither image set was uniquely associated with one method. This design helps limit repeated-image learning and stimulus-set confounding while preserving matched challenge types and correction objectives across conditions. However, because method order itself was not counterbalanced, general block-position effects, such as increasing familiarity with the study procedure or fatigue, cannot be separated from method effects.

\subsection{Participants}

30 participants completed the study, with paired survey data available for both correction methods. Participants were recruited from a university community, with no additional eligibility criteria applied. All study sessions were conducted in person and lasted approximately 40 minutes. Participants received \$20 in compensation for completing the study. The pre-study questionnaire recorded age group, gender, food-logging history and frequency, previous use of the manual-condition application, and nutrition-related education or work. Participants also rated, on 1--5 scales, their familiarity with manual food search/editing, AI photo-based logging, smartphone typing/searching, and portion estimation using common measures.

\subsection{Study Images and Reference Records}
\label{sec:study-images}

\begin{figure*}[t]
  \centering
  \mealstimulus{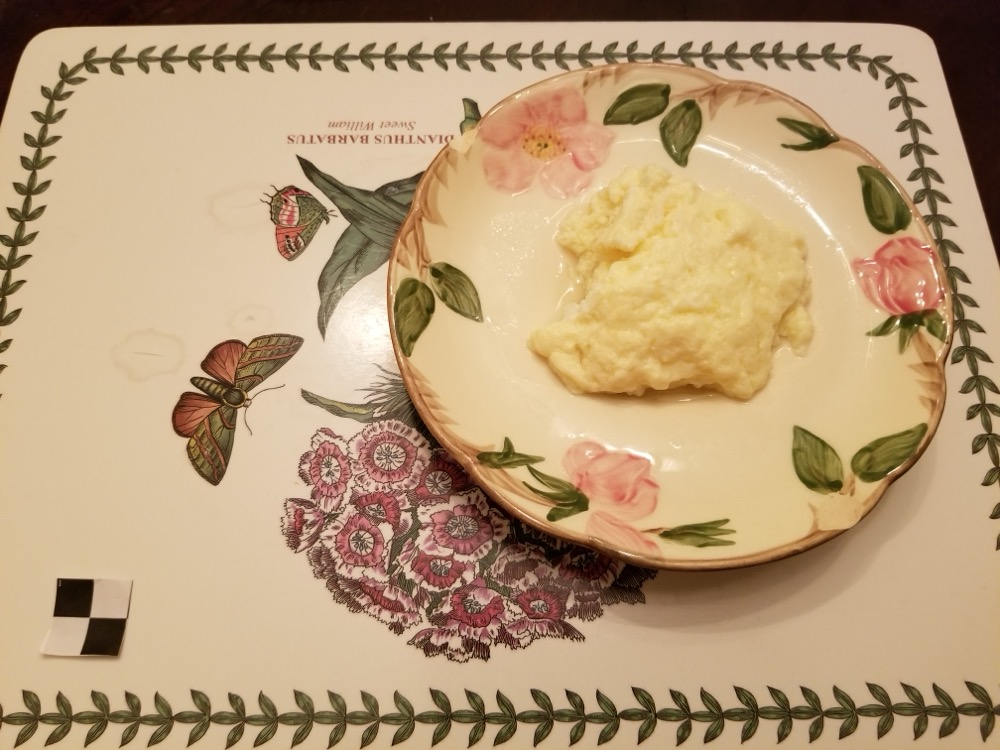}{A1: visual similarity}\hfill
  \mealstimulus{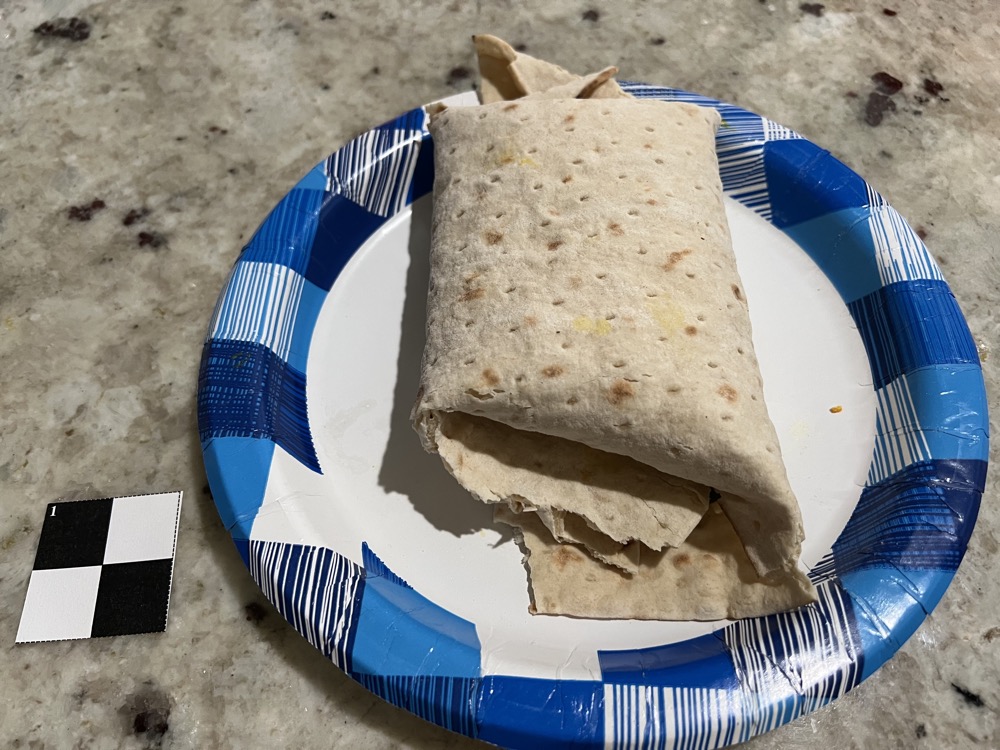}{A2: hidden composition}\hfill
  \mealstimulus{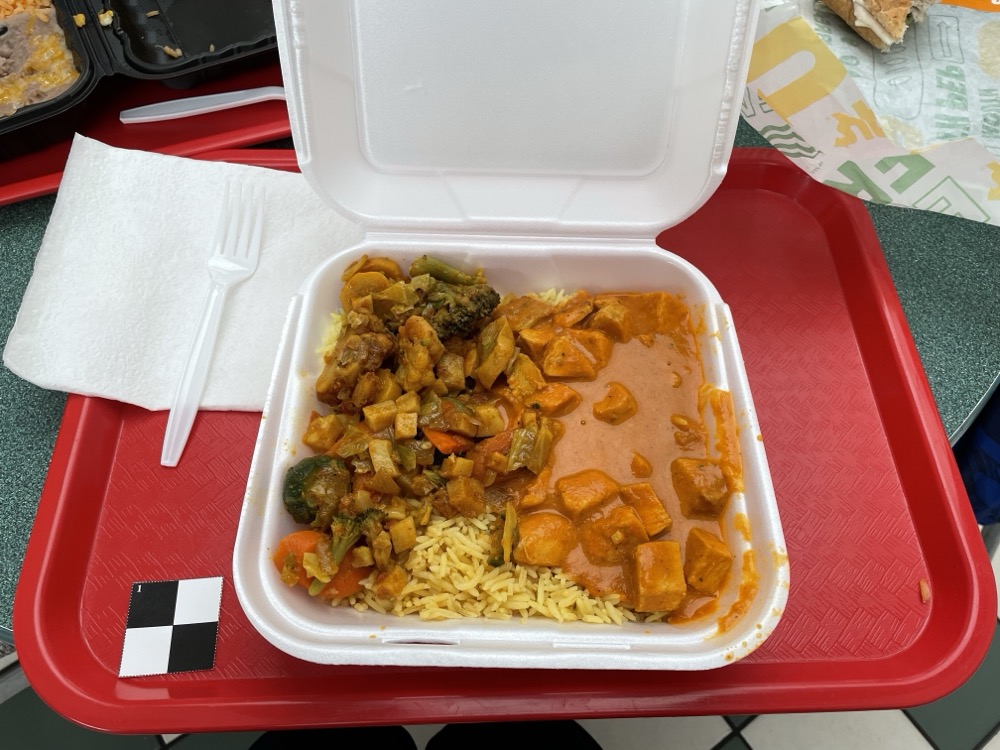}{A3: mixed dish}\hfill
  \mealstimulus[0 125bp 0 125bp]{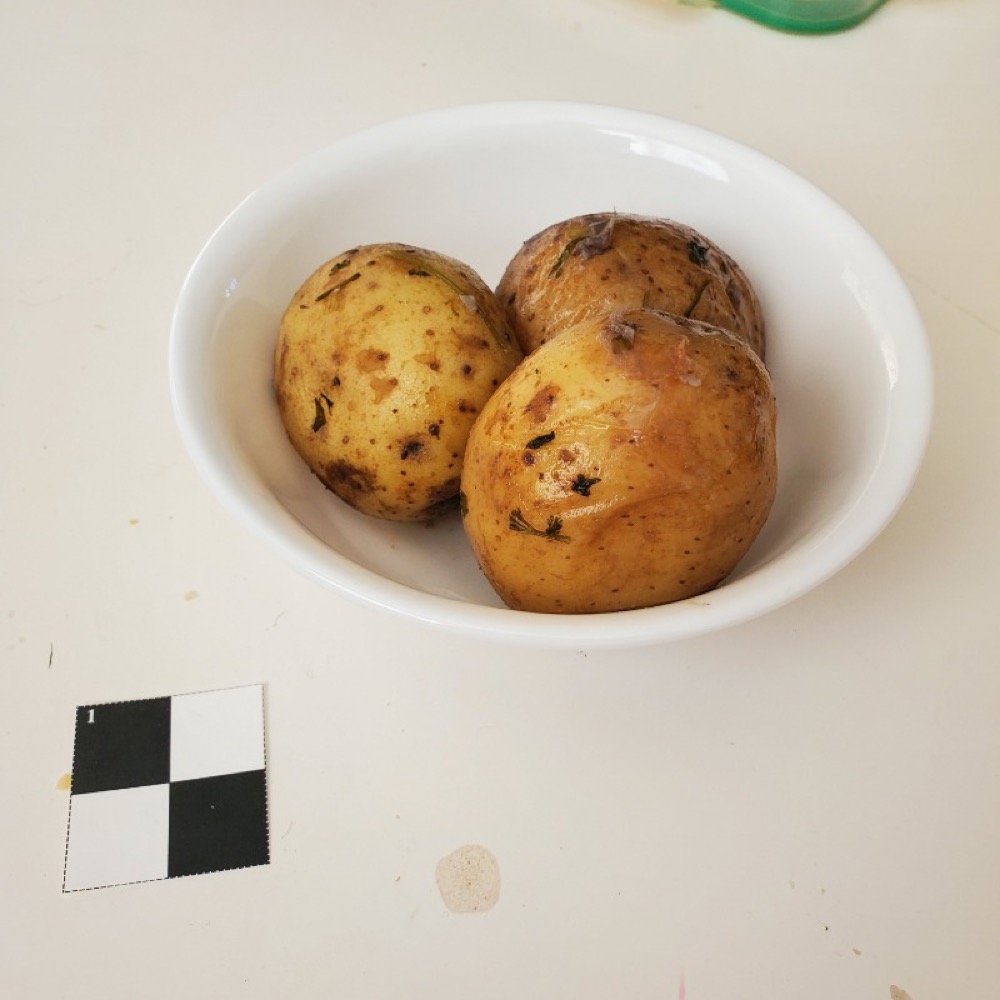}{A4: preparation}\hfill
  \mealstimulus{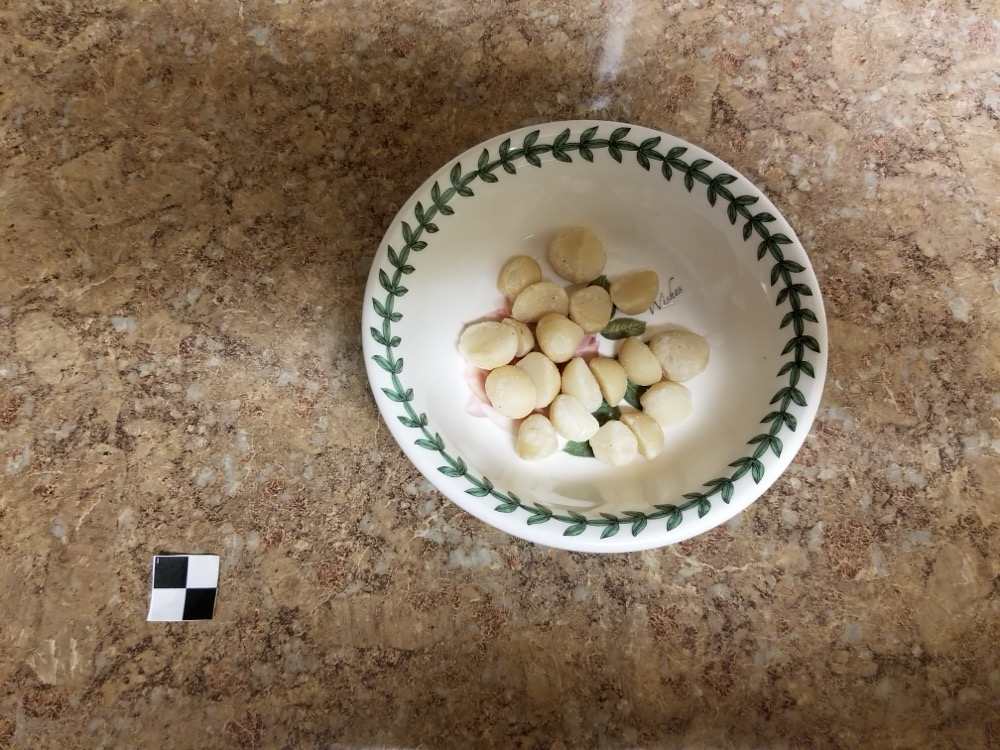}{A5: portion}

  \vspace{2mm}

  \mealstimulus[0 200bp 0 50bp]{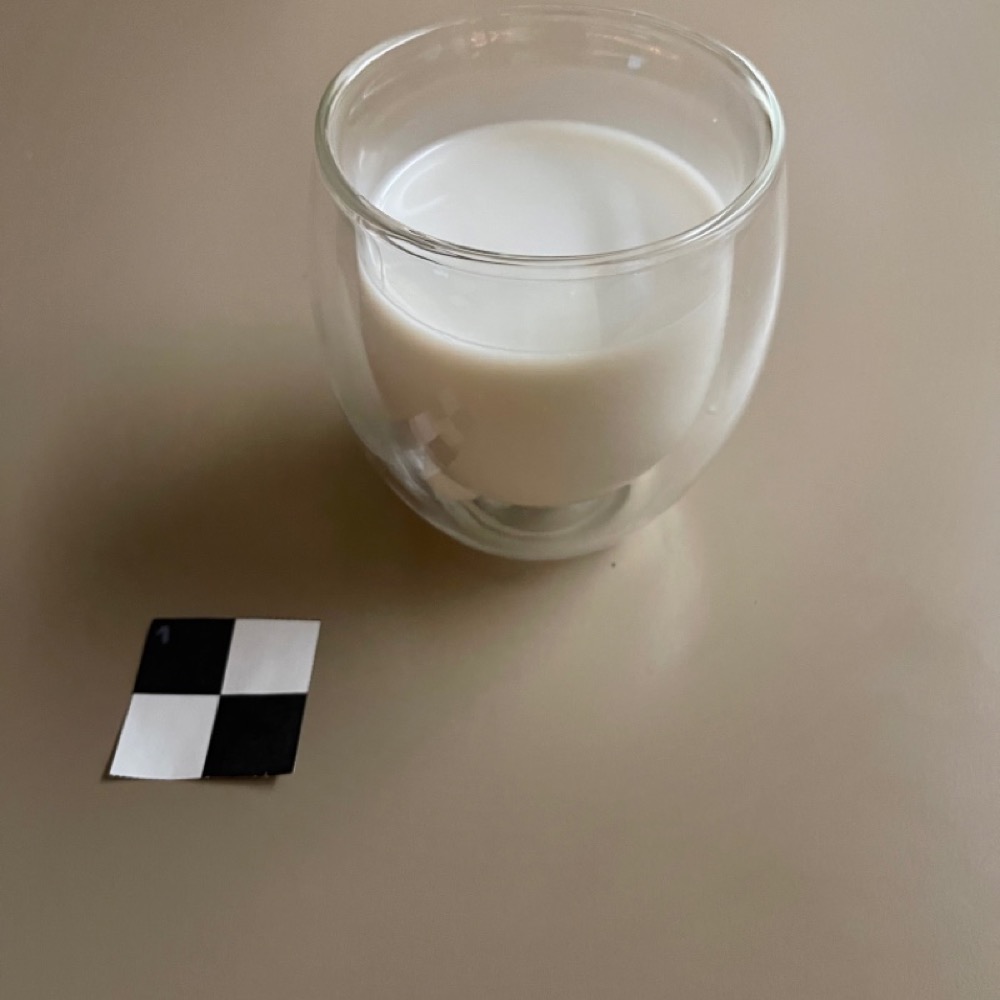}{B1: visual similarity}\hfill
  \mealstimulus[0 125bp 0 125bp]{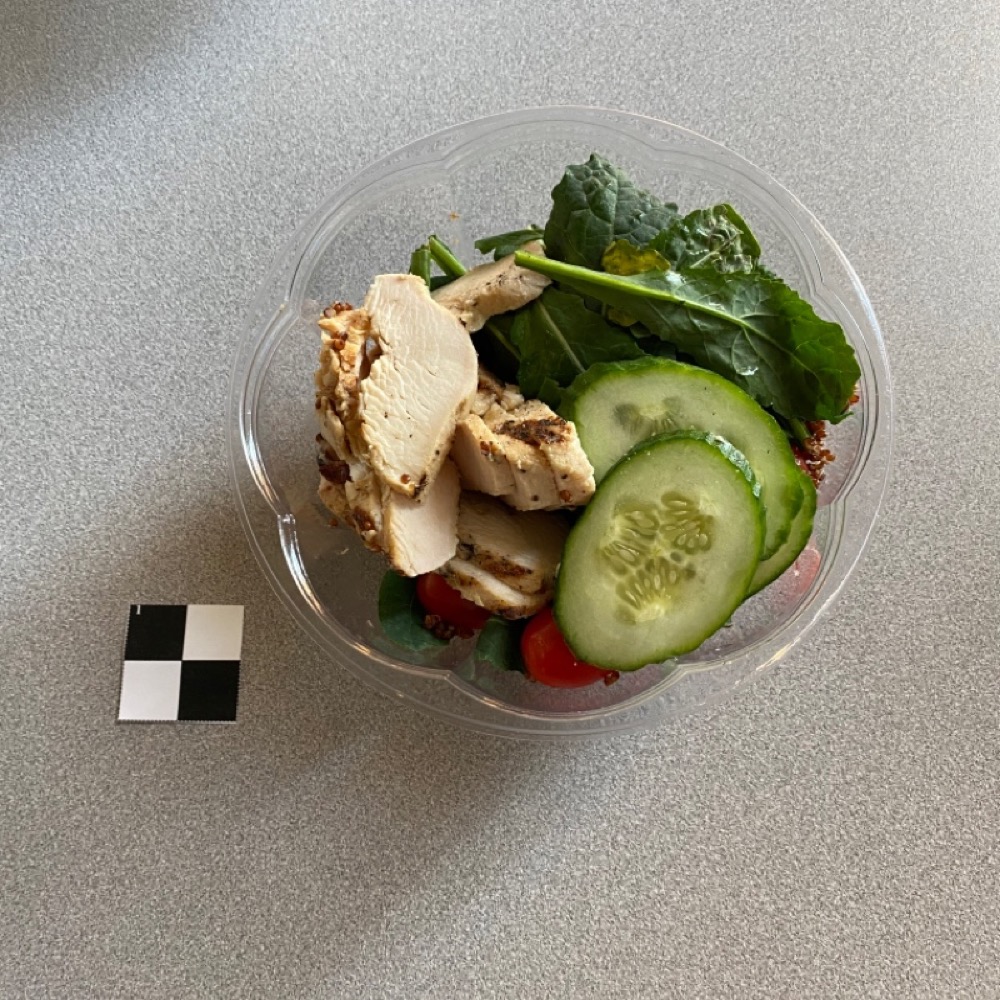}{B2: hidden composition}\hfill
  \mealstimulus[0 125bp 0 125bp]{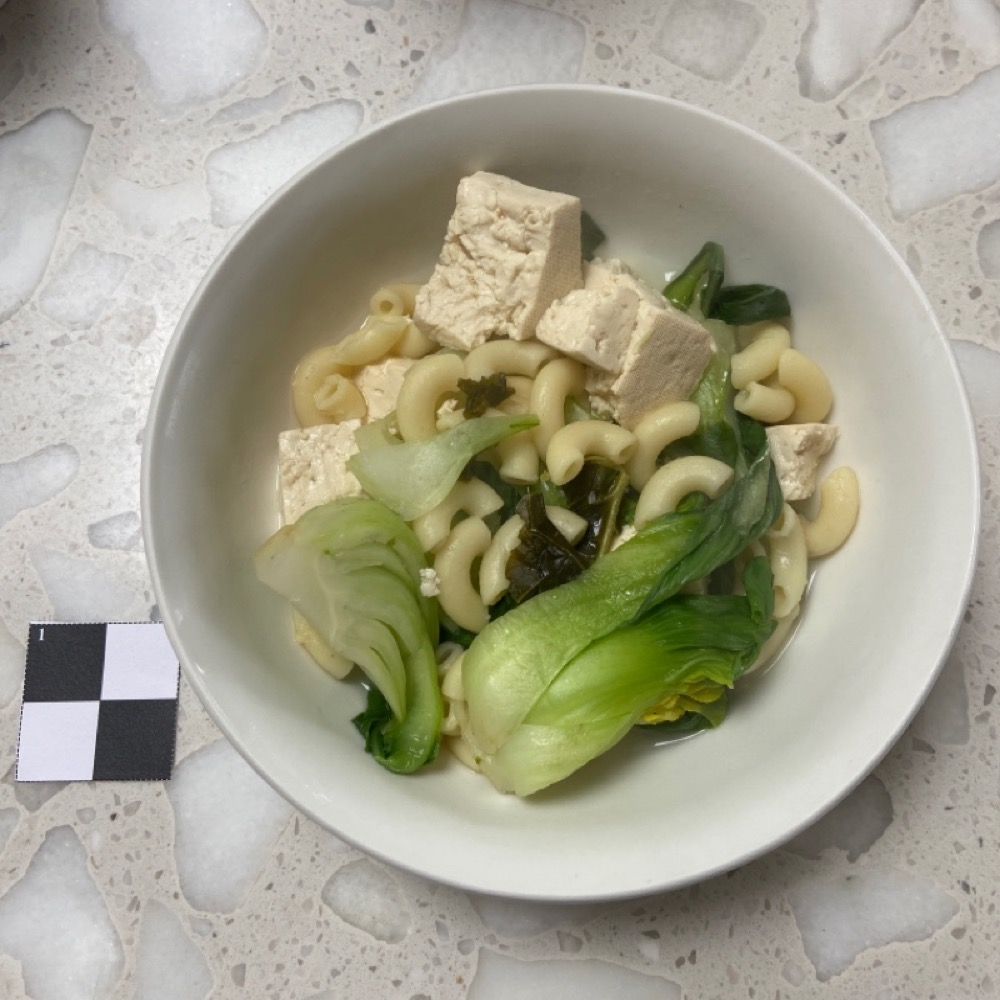}{B3: mixed dish}\hfill
  \mealstimulus[0 125bp 0 125bp]{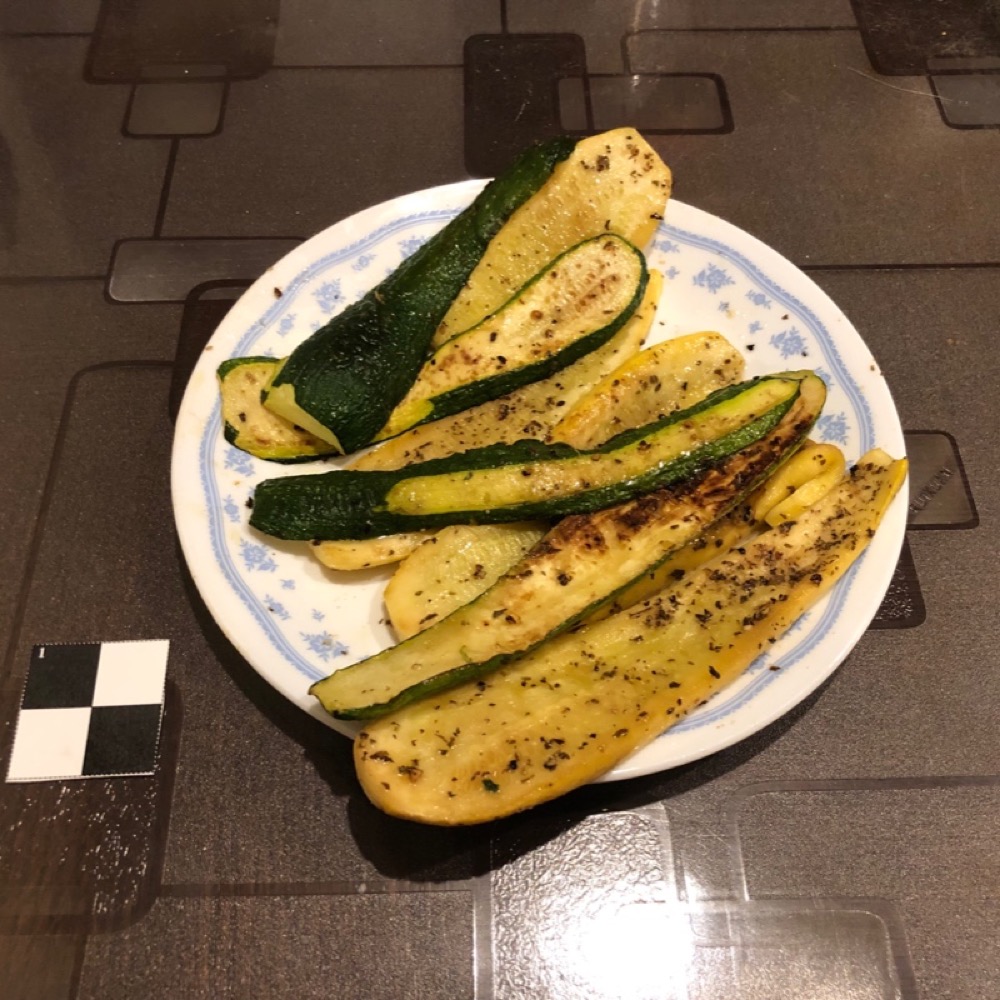}{B4: preparation}\hfill
  \mealstimulus[0 125bp 0 125bp]{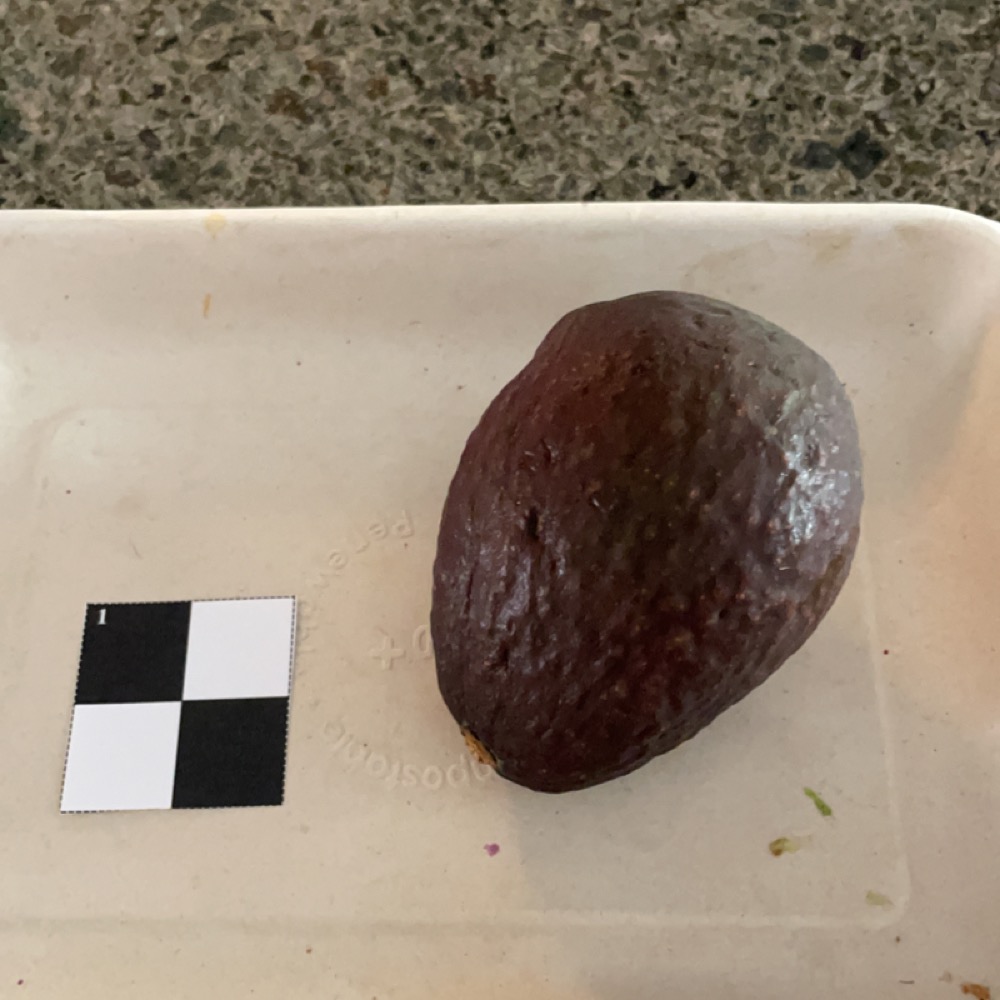}{B5: portion}

  \vspace{3mm}

  {\small
  \begin{tabular}{@{}lll@{}}
    \toprule
    Image-set allocation & Collaborative method & Manual method \\
    \midrule
    Allocation 1 & Set A (A1--A5) & Set B (B1--B5) \\
    Allocation 2 & Set B (B1--B5) & Set A (A1--A5) \\
    \bottomrule
  \end{tabular}\par}

  \vspace{1mm}
  \begin{minipage}{0.94\textwidth}
    \centering\small
    Task order within each method:
    1 $\rightarrow$ 2 $\rightarrow$ 3 $\rightarrow$ 4 $\rightarrow$ 5.\\
    Five images per method; ten distinct images per participant.
  \end{minipage}

  \caption{The ten selected SNAPMe images and image-set allocation. Each column
  contains two examples of the same assessment challenge. Sets A and B were
  rotated between correction methods across participants so that each
  participant completed five distinct images per method without repeating a
  photograph across methods.}
  \Description{Two rows of five photographs show paired examples of visual
  similarity, hidden composition, mixed dishes, preparation, and portion.
  An allocation matrix assigns Set A to collaborative correction and Set B to
  manual correction in one allocation and reverses the assignment in the
  other. Each participant completes five images with each correction method.}
  \label{fig:stimuli}
\end{figure*}

We screened photographs from Surveying Nutrient Assessment with Photographs of
Meals (SNAPMe), a dataset of 3,311 smartphone food photographs from real eating
occasions \cite{larke2023snapme}, to identify cases corresponding to the five
assessment challenges defined above. A key advantage of SNAPMe is that each
photograph is paired with an independently collected dietary record from the
National Cancer Institute's Automated Self-Administered 24-Hour Dietary
Assessment Tool (ASA24) \cite{larke2023snapme,nciasa24}, providing reference
labels for food identity, portion, and nutrient information that are independent
of the image. We selected two photographs per challenge, yielding ten study
images. Figure~\ref{fig:stimuli} shows the selected images and their organization
across the five challenges. Table~\ref{tab:stimuli} summarizes the corresponding
reference meals, portions, standardized starting errors, and correction targets.

\begin{table*}[t]
  \caption{Five assessment challenges and ten selected SNAPMe tasks. Reference
  portions were withheld from participants and used for accuracy assessment.
  The final column summarizes the standardized starting error and intended
  correction or verification.}
  \label{tab:stimuli}
  \footnotesize
  \begin{tabularx}{\textwidth}
    {@{}L{0.095\textwidth}L{0.045\textwidth}Y
    L{0.085\textwidth}L{0.405\textwidth}@{}}
    \toprule
    Challenge & Task & Reference food or meal & Reference portion &
    Starting error $\rightarrow$ correction target \\
    \midrule

    Visual similarity & A1 &
    Egg omelet or scrambled egg, made with cooking spray &
    100 g &
    Egg misidentified as mashed potatoes.
    $\rightarrow$ Egg made with cooking spray. \\

    Visual similarity & B1 &
    Almond milk, unsweetened &
    128.1 g &
    Almond milk misidentified as dairy milk.
    $\rightarrow$ Unsweetened almond milk. \\

    \addlinespace

    Hidden composition & A2 &
    Turkey luncheon meat, whole-wheat pita, raw spinach, American cheese,
    and mustard &
    253.95 g &
    Meal components omitted or bread misidentified.
    $\rightarrow$ Whole-wheat pita with all reference fillings. \\

    Hidden composition & B2 &
    Chicken breast, quinoa, cucumber, tomato, mixed salad greens, and olive oil &
    256.75 g &
    Quinoa and oil omitted.
    $\rightarrow$ Add missing ingredients and align remaining components with
    the reference. \\

    \addlinespace

    Mixed dish & A3 &
    Chicken breast, curry powder, cooked wild rice, mixed vegetables,
    and soy sauce &
    349.5 g &
    Incomplete or incorrect component representation.
    $\rightarrow$ Represent the reference meal components and judge portions. \\

    Mixed dish & B3 &
    Chinese cabbage, pasta, reduced-sodium chicken or turkey broth,
    tofu, and spinach &
    870 g &
    Meal components omitted or misidentified.
    $\rightarrow$ Correct the components and judge portions. \\

    \addlinespace

    Preparation & A4 &
    Potato, boiled, peel eaten, no added fat &
    510 g &
    Preparation requires correction or verification.
    $\rightarrow$ Boiled potato without added fat and image-judged portion. \\

    Preparation & B4 &
    Summer squash, cooked with oil &
    406 g &
    Added oil omitted.
    $\rightarrow$ Cooked-with-oil preparation and image-judged portion. \\

    \addlinespace

    Portion & A5 &
    Macadamia nuts &
    30 g &
    Incorrect portion.
    $\rightarrow$ Macadamia nuts with image-judged portion. \\

    Portion & B5 &
    Avocado, raw &
    150 g &
    Portion requires correction or verification.
    $\rightarrow$ Verify identity and judge portion. \\

    \bottomrule
  \end{tabularx}
\end{table*}

\subsection{Interaction Conditions and Procedure}

For \textbf{Manual Search-and-Edit Correction}, we used MyFitnessPal
(version 26.36.0) as a representative implementation of a conventional
database-based correction workflow \cite{evenepoel2020myfitnesspal}. The
comparison concerns the interaction method rather than the commercial product.
Participants inspected the existing record, searched for and selected food
entries, and manually adjusted servings, portions, or weights
within the MyFitnessPal interface. For
\textbf{Human--AI Collaborative Correction}, participants used the research
prototype described in Section~\ref{sec:system}. They could provide Text/Voice
Feedback, select Quick Clarification candidates, and use Visualized Portion Size
before submitting correction context for regeneration and verifying the returned
record. During onboarding, participants were encouraged to check Quick
Clarification first, use text or speech for unresolved corrections, and adjust
ingredient or portion information as needed.

All participants used the same study-provided iPhone 17 Pro Max with both
applications, study images, and starting records preloaded. Before the formal
tasks, the researcher introduced both correction methods and participants
practiced each using an image not included among the ten study images. Formal
trials began after participants reported familiarity with both workflows, ensuring that both interfaces had been introduced and practiced before formal measurement and thereby limiting first-exposure differences. Before
each trial, participants reviewed the meal photograph and food card and then
corrected the assigned starting record. A camera recorded the phone display and
touch activity throughout the correction task.

Each participant completed five tasks with each correction method, with a break
between the two blocks. After each block, participants completed Raw TLX and
method-specific experience questions. Condition-level ratings assessed ease,
confidence, perceived control, trust, and intended everyday use. A final
questionnaire compared the two methods and collected feature-specific ratings.

\subsection{Measures}
\label{sec:measures}

Measures correspond to the three parallel outcome domains in RQ1--RQ3.
Appendix~\ref{app:measurement-analysis} provides detailed scoring rules.

\subsubsection{Correction Burden (RQ1)}

Perceived correction burden was assessed using an adapted, unweighted NASA
Task Load Index (Raw TLX) \cite{hart1988nasatlx,hart2006tlx}. Raw TLX includes six dimensions---mental demand, physical demand,
temporal demand, performance, effort, and frustration---and is reported on a
0--100 scale, with higher scores indicating greater workload. The unweighted
composite requires all six dimensions, with performance oriented from
\emph{Perfect} to \emph{Failure}.

\subsubsection{Correction Efficiency and User Experience (RQ2)}

\textbf{Behavioral measures} comprised recorded usage time and click totals
for each participant and correction method across the five tasks. \textbf{User experience} was assessed using condition-level ratings of
confidence in the corrected record, perceived control, trust, and intended
everyday use.  Ease was rated from 1 (\emph{very difficult}) to 5 (\emph{very easy}). Overall preference, the method perceived as easier, and
method-specific feature ratings provided complementary descriptive measures of
participants' experience.

\subsubsection{Post-Correction Nutrient Accuracy (RQ3)}

Nutrient accuracy was assessed using absolute error in energy, protein,
carbohydrate, and fat relative to the reference values for each photograph.

\subsection{Data Analysis}
\label{sec:analysis}

Survey analyses use paired responses. We report
medians and quartiles and compare paired ratings using two-sided Wilcoxon
signed-rank tests \cite{wilcoxon1945ranking}, median paired differences, and
matched-pairs rank-biserial effects \cite{kerby2014difference} with bootstrap
95\% confidence intervals \cite{efron1994bootstrap}. The Raw TLX composite
retains its unadjusted test. Holm adjustment \cite{holm1979multiple} is applied
separately across the six Raw TLX dimensions and across ease, confidence,
perceived control, trust, and intended use. Preference, perceived easier
method, and method-specific feature ratings are reported descriptively.

Behavioral analyses use available participant pairs. Total recorded usage time and
click totals are compared using two-sided Wilcoxon signed-rank tests, with Holm
adjustment across the two outcomes. We report descriptive statistics, paired
differences, and rank-biserial effects with bootstrap 95\% confidence intervals.
For nutrient accuracy, entry-level absolute errors are averaged within each
photograph and method and then equally across the ten photographs. Method
differences are tested using exact two-sided paired sign-flip tests
\cite{winkler2014permutation}, with Holm adjustment across energy, protein,
carbohydrate, and fat.

\subsection{Ethics and Data Management}

The study received approval from the relevant institutional ethics review board
(details withheld for anonymized review). Participants consented to the collection
and reporting of their demographic and background information and to the recording
of their hand gestures during study tasks. Study data included interaction
recordings, application logs, corrected dietary records, and questionnaires.
Recordings are kept confidential and will not be shared without additional consent
from the participants. De-identified survey data may be made available upon request.


\section{Results}
\label{sec:results}

\subsection{Participants}

As summarized in Table~\ref{tab:participants}, all 30 participants contributed paired survey data for both correction methods. Most participants were aged 25--34 (66.7\%), and 19 (63.3\%) were men. More than half (53.3\%) had never previously logged their diet, and 73.3\% had never used MyFitnessPal. Familiarity with manual food search/editing and AI photo-based logging was generally low, whereas comfort with smartphone typing and searching was high.

\begin{table}[t]
  \caption{Participant characteristics ($N=30$). Categorical values are $n$ (\%); familiarity and comfort ratings are
  medians [Q1, Q3] on 1--5 scales.}
  \label{tab:participants}
  \small
  \begin{tabularx}{\linewidth}{@{}L{0.43\linewidth}Y@{}}
    \toprule
    Characteristic & Final value \\
    \midrule
    Participants contributing study data & 30 \\
    Age group, $n$ (\%) & 18--24: 8 (26.7); 25--34: 20 (66.7); 35--44: 2 (6.7) \\
    Gender, $n$ (\%) & Men: 19 (63.3); women: 11 (36.7) \\
    Food-logging history, $n$ (\%) & Never: 16 (53.3); previously: 12 (40.0); currently: 2 (6.7) \\
    Current logging frequency, $n$ (\%) & Daily: 4 (13.3); a few times a week: 2 (6.7); a few times a month: 3 (10.0); rarely: 8 (26.7); never: 13 (43.3) \\
    Previous use of MyFitnessPal, $n$ (\%) & Previously used: 8 (26.7); never used: 22 (73.3) \\
    Nutrition education/work experience, $n$ (\%) & Yes: 7 (23.3); no: 23 (76.7) \\
    Manual search/edit familiarity, median [Q1, Q3] & 2 [1, 3] \\
    AI-photo-logging familiarity, median [Q1, Q3] & 2 [1, 3] \\
    Smartphone typing/search comfort, median [Q1, Q3] & 5 [4, 5] \\
    Portion-estimation familiarity, median [Q1, Q3] & 3 [2, 3] \\
    \bottomrule
  \end{tabularx}
\end{table}

\subsection{RQ1: Correction Burden}
\label{sec:results-rq1}

\begin{figure*}[p]
  \centering
  \includegraphics[width=0.975\textwidth]{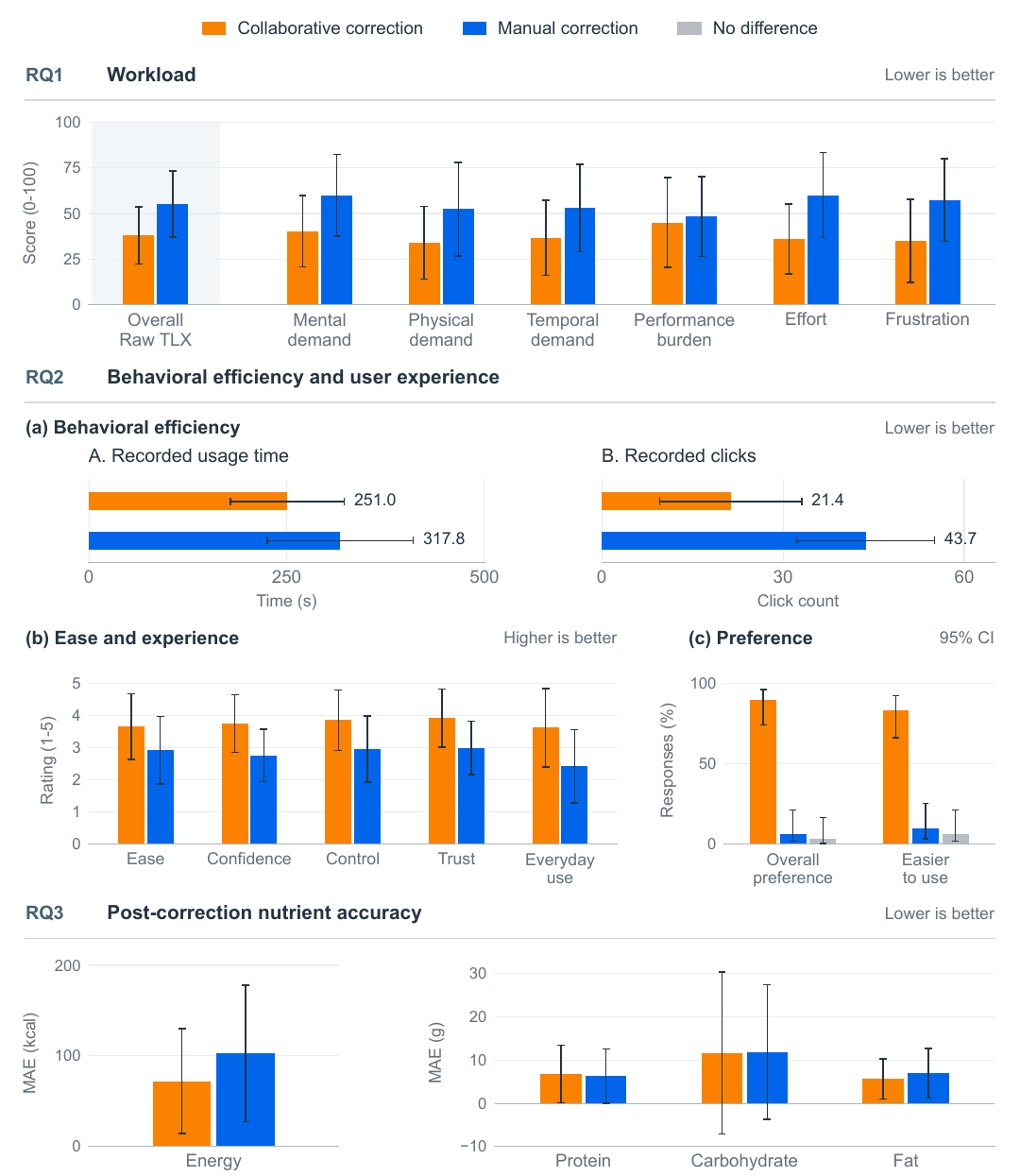}
  \captionsetup{skip=4pt}
  \caption{Collaborative (orange) and manual (blue) correction across the three research questions. RQ1 shows Raw TLX and its six dimensions on a 100-point scale, with lower values indicating lower workload. RQ2 shows (a) recorded usage time and click totals, with lower values indicating greater behavioral efficiency; (b) ease, confidence, perceived control, trust, and intended everyday use on 1--5 scales, with higher values indicating more favorable ratings; and (c) overall preference and the method perceived as easier to use. RQ3 shows energy and macronutrient mean absolute errors (MAEs), with lower values indicating closer nutrient agreement with the reference. Bars show means
  with $\pm1$ sample SD across participants, or across photo-level means for RQ3; preference bars show percentages with pointwise Wilson 95\% confidence intervals.}
  \Description{The figure compares collaborative and manual correction across perceived workload, behavioral efficiency, user experience, preference, and nutrient accuracy. Collaborative correction shows lower workload and more favorable user-experience ratings. The behavioral panel compares recorded usage time and click totals. The accuracy panel compares energy, protein, carbohydrate, and fat mean absolute errors across ten study photographs.}
  \label{fig:combined-results}
\end{figure*}

As summarized in Table~\ref{tab:paired-survey} and shown in the RQ1 panel of Figure~\ref{fig:combined-results}, collaborative correction was associated with substantially lower perceived workload than manual correction. Median Raw TLX was 35.83 [25.83, 47.50] for collaborative correction and 55.00 [45.83, 60.00] for manual correction, with a median paired difference of $-12.50$ points (95\% CI [$-16.67$, $-10.00$]), $r_{\mathrm{rb}}=-.78$ (95\% CI [$-.99$, $-.49$]), and $p<.001$. As summarized in Table~\ref{tab:tlx-dimensions}, five of the six workload dimensions were significantly lower with collaborative correction after Holm adjustment: mental demand and physical demand (both $p_{\mathrm{Holm}}=.0052$), temporal demand ($p_{\mathrm{Holm}}=.0103$), and effort and frustration (both $p_{\mathrm{Holm}}=.0011$). Median paired differences were $-15$, $-15$, $-10$, $-20$, and $-20$ points, respectively, while perceived performance did not differ significantly ($r_{\mathrm{rb}}=-.18$, $p_{\mathrm{Holm}}=.4981$).

\begin{table*}[t]
  \caption{Paired condition-level survey outcomes. Values are medians [Q1, Q3]. $\Delta$ is Collaborative minus Manual, with paired bootstrap 95\% CI. $r_{\mathrm{rb}}$ is the matched-pairs rank-biserial effect with bootstrap 95\% CI. Raw TLX retains its unadjusted $p$ value; the five experience ratings use Holm adjustment as one comparison family.}
  \label{tab:paired-survey}
  \footnotesize
  \begin{tabularx}{\textwidth}{@{}Y L{0.14\textwidth}L{0.14\textwidth}L{0.15\textwidth}L{0.16\textwidth}L{0.075\textwidth}@{}}
    \toprule
    Outcome & Collaborative & Manual & Median $\Delta$ [95\% CI] & $r_{\mathrm{rb}}$ [95\% CI] & $p$ \\
    \midrule
    Raw TLX workload & \shortstack[l]{35.83\\{[25.83, 47.50]}} & \shortstack[l]{55.00\\{[45.83, 60.00]}} & \shortstack[l]{-12.50\\{[-16.67, -10.00]}} & \shortstack[l]{-0.78\\{[-0.99, -0.49]}} & $<.001$ \\
    Ease, 1--5 & \shortstack[l]{4.00\\{[3.00, 4.00]}} & \shortstack[l]{3.00\\{[2.00, 4.00]}} & \shortstack[l]{1.00\\{[0.00, 1.00]}} & \shortstack[l]{0.54\\{[0.13, 0.89]}} & $.0203$ \\
    Confidence, 1--5 & \shortstack[l]{4.00\\{[3.00, 4.00]}} & \shortstack[l]{3.00\\{[2.00, 3.00]}} & \shortstack[l]{1.00\\{[1.00, 1.00]}} & \shortstack[l]{0.89\\{[0.70, 1.00]}} & $<.001$ \\
    Perceived control, 1--5 & \shortstack[l]{4.00\\{[3.25, 4.00]}} & \shortstack[l]{3.00\\{[2.00, 4.00]}} & \shortstack[l]{1.00\\{[0.00, 1.50]}} & \shortstack[l]{0.72\\{[0.37, 1.00]}} & $.0052$ \\
    Trust, 1--5 & \shortstack[l]{4.00\\{[4.00, 4.75]}} & \shortstack[l]{3.00\\{[2.25, 4.00]}} & \shortstack[l]{1.00\\{[0.00, 1.00]}} & \shortstack[l]{0.93\\{[0.76, 1.00]}} & $<.001$ \\
    Intended everyday use, 1--5 & \shortstack[l]{4.00\\{[3.00, 4.75]}} & \shortstack[l]{3.00\\{[1.00, 3.00]}} & \shortstack[l]{1.00\\{[0.50, 1.50]}} & \shortstack[l]{0.95\\{[0.80, 1.00]}} & $<.001$ \\
    \bottomrule
  \end{tabularx}
\end{table*}

\begin{table}[t]
  \caption{Raw TLX dimension results. Values are medians [Q1, Q3] on a 100-point scale. Effects are matched-pairs rank-biserial $r_{\mathrm{rb}}$ [bootstrap 95\% CI] for Collaborative minus Manual; negative values favor collaborative correction. $p$ values use Holm adjustment across the six dimensions.}
  \label{tab:tlx-dimensions}
  \scriptsize
  \begin{tabularx}{\linewidth}{@{}Y L{0.19\linewidth}L{0.19\linewidth}L{0.22\linewidth}L{0.11\linewidth}@{}}
    \toprule
    Dimension & Collaborative & Manual & $r_{\mathrm{rb}}$ [95\% CI] & $p_{\mathrm{Holm}}$ \\
    \midrule
    Mental & \shortstack[l]{40.00\\{[30.00, 50.00]}} & \shortstack[l]{60.00\\{[50.00, 70.00]}} & \shortstack[l]{-0.77\\{[-0.98, -0.43]}} & $.0052$ \\
    Physical & \shortstack[l]{30.00\\{[20.00, 40.00]}} & \shortstack[l]{50.00\\{[30.00, 70.00]}} & \shortstack[l]{-0.80\\{[-0.97, -0.49]}} & $.0052$ \\
    Temporal & \shortstack[l]{35.00\\{[20.00, 50.00]}} & \shortstack[l]{50.00\\{[40.00, 67.50]}} & \shortstack[l]{-0.69\\{[-1.00, -0.28]}} & $.0103$ \\
    Performance & \shortstack[l]{40.00\\{[20.00, 60.00]}} & \shortstack[l]{50.00\\{[30.00, 60.00]}} & \shortstack[l]{-0.18\\{[-0.69, 0.36]}} & $.4981$ \\
    Effort & \shortstack[l]{30.00\\{[20.00, 50.00]}} & \shortstack[l]{60.00\\{[50.00, 80.00]}} & \shortstack[l]{-0.85\\{[-1.00, -0.61]}} & $.0011$ \\
    Frustration & \shortstack[l]{30.00\\{[20.00, 50.00]}} & \shortstack[l]{60.00\\{[42.50, 70.00]}} & \shortstack[l]{-0.87\\{[-1.00, -0.62]}} & $.0011$ \\
    \bottomrule
  \end{tabularx}
\end{table}

\subsection{RQ2: Correction Efficiency and User Experience}
\label{sec:results-rq2}

\subsubsection{RQ2a: Behavioral Efficiency}

As shown in the RQ2(a) panel of Figure~\ref{fig:combined-results}, collaborative correction had lower recorded usage time and fewer recorded clicks than manual correction. The aggregate analysis used available paired records covering the five tasks completed with each method. Mean recorded usage time was 251.05~s (SD 71.85) for collaborative correction and 317.77~s (SD 92.33) for manual correction; the collaborative-condition mean was 21.0\% lower than the manual-condition mean. Median recorded usage times were 254.00 [205.50, 278.50] and 283.50 [250.00, 375.75]~s, respectively. The median paired difference was $-57.50$~s (95\% CI [$-89.00$, $-25.50$]), with
$W=10$, $r_{\mathrm{rb}}=-.92$ (95\% CI [$-1.00$, $-.72$]), and $p_{\mathrm{Holm}}=.00016$. Collaborative correction also had fewer recorded clicks, with means of 21.36 (SD 11.76) versus 43.68 (SD 11.41); the collaborative-condition mean was 51.1\% lower than the manual-condition mean. Median recorded click totals were 17.00 [14.25, 24.50] and 44.00 [34.25, 49.00], with a median paired difference of $-21.50$ clicks (95\% CI [$-27.00$, $-17.50$]), $W=0$, $r_{\mathrm{rb}}=-1.00$ (95\% CI [$-1.00$, $-1.00$]), and $p_{\mathrm{Holm}}=.00008$.

\subsubsection{RQ2b: User Experience}

As summarized in Table~\ref{tab:paired-survey} and shown in the RQ2(b--c) panels of Figure~\ref{fig:combined-results}, collaborative correction received more favorable user-experience ratings and was strongly preferred over manual
correction. Median ease ratings were 4 [3, 4] and 3 [2, 4], respectively, with a median paired difference of 1 point (95\% CI [0, 1]), $r_{\mathrm{rb}}=.54$ (95\% CI [.13, .89]), and $p_{\mathrm{Holm}}=.0203$. Confidence, perceived control, trust, and intended everyday use each had a median of 4 for collaborative correction and 3 for manual correction, with median paired differences of 1 point. Rank-biserial effects were .89 for
confidence, .72 for perceived control, .93 for trust, and .95 for intended use. After Holm adjustment, all four differences remained significant: $p_{\mathrm{Holm}}<.001$ for confidence, trust, and intended everyday use, and
$p_{\mathrm{Holm}}=.0052$ for perceived control. Overall, 27 participants (90.0\%) preferred collaborative correction, compared with two (6.7\%) who preferred manual correction and one (3.3\%) who reported no difference.
Twenty-five participants (83.3\%) also identified collaborative correction as easier to use, compared with three (10.0\%) who selected manual correction and two (6.7\%) who reported no difference. The Wilson 95\% confidence intervals for the collaborative shares were [74.4\%, 96.5\%] for overall preference and [66.4\%, 92.7\%] for perceived ease.

As summarized in Table~\ref{tab:feature-ratings}, method-specific ratings further characterized participants' experience with the two workflows. Manual food-search ease and portion-entry ease each had a median of 3 [2, 3], while
step manageability had a median of 2.5 [2, 3]. All collaborative-correction items had a median of 4. Quick Clarification helpfulness was rated 4 or 5 by 22 participants (73.3\%), and Visualized Portion Size helpfulness by 21 participants (70.0\%). Because the feature questions differed between the two methods, these ratings are reported descriptively rather than as paired comparisons.

\begin{table}[t]
  \caption{Descriptive method-specific ratings. Values are medians [Q1, Q3] on 1--5 scales. Because the items differed between methods, they are reported descriptively rather than as paired comparisons.}
  \label{tab:feature-ratings}
  \scriptsize
  \begin{tabularx}{\linewidth}{@{}Y L{0.25\linewidth}@{}}
    \toprule
    Item & Median [Q1, Q3] \\
    \midrule
    \multicolumn{2}{@{}l}{\textit{Manual correction}} \\
    Food-search ease & 3 [2, 3] \\
    Portion-entry ease & 3 [2, 3] \\
    Step manageability & 2.5 [2, 3] \\
    \addlinespace
    \multicolumn{2}{@{}l}{\textit{Collaborative correction}} \\
    Method behaved as expected & 4 [3, 4] \\
    Quick Clarification helpfulness & 4 [3.25, 4.75] \\
    Visualized Portion Size helpfulness & 4 [3, 5] \\
    Quick Clarification relevance & 4 [4, 4] \\
    Quick Clarification made correction easier & 4 [3.25, 4.75] \\
    Appropriate Quick Clarification option available & 4 [3.25, 4] \\
    Quick Clarification option-list manageability & 4 [4, 4.75] \\
    Visualized Portion Size clarity & 4 [4, 5] \\
    Visualized Portion Size made portion correction easier & 4 [4, 5] \\
    \bottomrule
  \end{tabularx}
\end{table}

\subsection{RQ3: Post-Correction Nutrient Accuracy}
\label{sec:results-rq3}

As shown in the RQ3 panel of Figure~\ref{fig:combined-results}, in this exploratory photo-level analysis, the collaborative workflow had lower energy mean absolute error (MAE) than the manual workflow across the ten study photographs: 72.12 versus 102.84~kcal, corresponding to a 29.9\% lower MAE ($p_{\mathrm{Holm}}=.031$). Protein MAE was 6.78 versus 6.32~g, carbohydrate MAE was 11.69 versus 11.90~g, and fat MAE was 5.66 versus 6.96~g for collaborative and manual correction, respectively; the differences in protein, carbohydrate, and fat MAE were not statistically significant after Holm adjustment (all $p_{\mathrm{Holm}}=1.00$). Overall, the collaborative workflow showed lower observed energy error, while macronutrient errors remained in a similar range to those of the representative commercial manual workflow. These findings suggest that the lower observed interaction burden was not accompanied by an obvious deterioration in nutrient agreement. Because the two conditions combined different correction interfaces with different underlying dietary-assessment systems---DietAI24 for collaborative correction and MyFitnessPal for manual correction---the nutrient findings should be interpreted as an end-to-end comparison of the two workflows rather than as the isolated effect of the correction method itself.


\section{Discussion}
\label{sec:discussion}

In this within-participant comparison, Human–AI Collaborative Correction was associated with lower experienced and behavioral correction work than Manual Search-and-Edit Correction. Collaborative correction was associated with lower perceived workload, lower recorded usage time, fewer recorded clicks, and more favorable ratings of ease, confidence, perceived control, trust, and intended everyday use; 90.0\% of participants preferred it overall. The collaborative workflow also showed lower energy error, while protein, carbohydrate, and fat errors remained in a similar observed range to the representative commercial manual workflow. Together, these findings support intent-driven regeneration as a promising interaction approach for reducing the work involved in correcting structured dietary records while retaining user verification of the regenerated output.

\subsection{Redistributing Correction Work}

The central interaction change in collaborative correction is not simply the use of AI, but a redistribution of correction work between the user and the system. Manual correction requires users to translate meal knowledge into database searches, entry selection, and field-level edits. Collaborative correction instead allows users to communicate correction intent while the system performs the corresponding updates to linked food, portion, weight, and nutrient fields. The lower Raw TLX workload, lower recorded usage time, and fewer recorded clicks provide converging subjective and behavioral evidence consistent with less work required to implement corrections under the collaborative workflow. The reductions in mental, physical, and temporal demand, effort, and frustration further suggest that the benefit extended beyond simply reducing the number of interface actions.

This finding provides a concrete instance of Hutchins et al.'s distinction between the gulfs of execution and evaluation \cite{hutchins1985direct}. Conceptually, collaborative correction shifts execution work away from translating an intended change into low-level database operations, but it does not eliminate evaluation work: users must still determine whether the regenerated record
reflects their intent. In this sense, our work extends the principle of semi-automated tracking from data capture to correction. Prior work has emphasized complementary roles for automation and manual contribution in self-monitoring \cite{choe2017semiautomated}; here, the user contributes meal knowledge while the system performs the representational work that prior food-journaling studies have identified as burdensome \cite{cordeiro2015barriers,luo2019codesigning,silva2025foodytalk}.

The behavioral findings further show that reducing low-level interaction does not translate proportionally into elapsed-time savings. Recorded clicks were 51.1\% lower in the collaborative condition, compared with a 21.0\% lower mean recorded usage time. The aggregate measures do not distinguish time spent formulating correction intent, waiting for system processing, and inspecting the regenerated record. Thus, efficiency should not be understood only as minimizing elapsed time: reducing repeated search, selection, and editing operations can meaningfully decrease interaction work even when regeneration and verification remain part of the workflow.

\subsection{Regeneration, Verification, and Output Quality}

Delegating field-level updates changes the form of user control. Manual correction provides direct operational control over individual fields, whereas collaborative correction provides control over the intended change, regeneration, and acceptance. Participants nevertheless reported higher perceived control with collaborative correction. This suggests that a sense of control does not necessarily require users to execute every low-level operation themselves. By separating selection of correction intent from regeneration and returning the regenerated record for inspection before acceptance, the workflow places explicit user authorization both before and after AI action. This provides a specific instance of the direct-manipulation/delegation tradeoff \cite{shneiderman1997direct,amershi2019guidelines}.

Verification is particularly important because dietary-record fields are interdependent. Changing food identity can alter the applicable portion representation, gram weight, and derived nutrient estimates, so a semantic correction may propagate beyond the field the user initially intended to change. The favorable confidence and trust ratings therefore need to be interpreted alongside objective output quality. Trust should be calibrated to system capabilities \cite{leesee2004trust}, and prior work has shown that favorable subjective judgments do not necessarily imply correct AI-assisted decisions \cite{bucinca2021trust}. In our study, the collaborative workflow achieved 29.9\% lower energy MAE, while the observed macronutrient errors were of similar magnitude to those of the manual workflow. This is encouraging because the reductions in workload and interaction effort were not accompanied by an obvious deterioration in the observed nutrient outcomes.

The nutrient-accuracy findings should nevertheless be interpreted at the level of the complete workflows. Collaborative correction used DietAI24 to regenerate the dietary record, whereas manual correction was implemented through MyFitnessPal. The observed nutrient errors therefore reflect both the interaction method and the underlying dietary-assessment system rather than isolating the effect of
correction method alone. From an end-to-end perspective, however, the finding remains important: the collaborative workflow substantially reduced correction burden and interaction effort while achieving lower energy error and macronutrient errors in a similar observed range to the representative commercial manual workflow.

\subsection{Supporting Different Ways to Express Correction Intent}

Human--AI Collaborative Correction combines three complementary ways of communicating correction intent rather than relying on free-form prompting alone. Text/Voice Feedback allows users to state what they know in their own words; Quick Clarification converts model-based auditing into plausible alternatives that users can recognize and select; and Visualized Portion Size
provides a visual way to communicate intended amount. Participants responded favorably to these mechanisms: all collaborative feature items had a median of 4, while 73.3\% rated Quick Clarification helpfulness and 70.0\% rated Visualized Portion Size helpfulness as 4 or 5. These findings are consistent with prior work showing both the difficulty of formulating prompts and the value of structured refinement controls \cite{subramonyam2024gulf,wang2024promptcharm}.

The five dietary-assessment challenges help explain why complementary interaction mechanisms may be useful. Visually similar foods may require correction of food identity; hidden ingredients and preparation details may require contextual information known only to the eater; mixed dishes may require several coordinated changes; and portion correction requires translating a
user's judgment of amount into a structured representation. Quick Clarification supports recognition of plausible alternatives, Text/Voice Feedback provides a path for information not captured by predefined choices, and Visualized Portion Size supports communication of amount without requiring direct translation into grams or database units. Because these mechanisms were evaluated together as one workflow, the present study supports the combined interaction design rather than attributing the observed benefits to any single component.

\subsection{Design Implications and Broader Applicability}

Our findings suggest several implications for AI-assisted correction interfaces. First, correction can be organized around \emph{semantic intent rather than database operations}. When users already know what is wrong, interfaces can allow them to communicate that knowledge directly and delegate its translation into linked structured fields to the system. Second, such delegation should be paired with \emph{inspectable regeneration}. Because one semantic correction can propagate across multiple fields, users need clear opportunities to examine the resulting changes before accepting them. Future interfaces could further support this work through changed-field summaries, highlighting of propagated updates, or selective acceptance of individual changes, consistent with prior work on transparent and controllable human--AI interaction \cite{amershi2019guidelines,wu2022aichains}.

Third, designers should consider interaction effort and system latency together. The larger reduction in recorded clicks than in recorded usage time indicates that removing low-level operations does not automatically remove all elapsed time. Faster regeneration may therefore increase the practical benefit of intent-driven correction. At the same time, direct manipulation may remain useful when users know an exact value or want to make a localized change. Hybrid interfaces that allow users to move flexibly between semantic regeneration and direct editing may therefore be valuable \cite{suhm2001multimodal,masson2024directgpt,shneiderman1997direct}. More broadly, correction systems should be evaluated across burden, behavioral efficiency, user experience, and output quality together: an accurate workflow may still be impractical if repeated correction is burdensome, whereas a low-effort workflow is insufficient if it produces inaccurate records.

Human--AI Collaborative Correction may also provide a useful interaction pattern for AI-generated structured records beyond dietary assessment. The pattern is most relevant when users possess contextual information unavailable to the AI and a semantic correction can affect multiple interdependent fields. Examples include calendar records generated from messages or expense records extracted
from receipts, where users may be able to state what is wrong more easily than specify every field-level operation needed to repair it. Unlike interactive machine learning, where correction may become a training signal \cite{amershi2014power}, the present workflow uses correction context to refine the current record without updating the underlying model. Whether this allocation of intent communication, regeneration, and verification benefits other structured-record domains remains an empirical question.

\section{Limitations and Future Work}
\label{sec:limitations}

Several limitations qualify these findings. First, the study was designed to compare how users implement known corrections rather than how they detect errors during everyday dietary tracking. Food identity, composition, and preparation were provided through food cards, while participants judged portion from the photographs. The ten SNAPMe images were purposively selected to represent five recurring challenges and therefore do not capture the prevalence or full diversity of errors encountered in real-world dietary assessment. With only two images per challenge, the study also does not support strong conclusions about differences among challenge types. Future field studies should evaluate the complete workflow from meal capture and error recognition through correction and verification during participants' own dietary logging.

Second, method order was fixed, with manual correction preceding collaborative correction for all participants. We sought to limit stimulus-specific carryover by using non-overlapping photographs across methods within each participant and rotating the two matched image sets between methods across participants. Both workflows were also introduced and practiced before formal measurement. These controls help limit repeated-image learning, stimulus-set confounding, and differences attributable to first exposure to an interface. However, they cannot eliminate general block-order effects, such as increased familiarity with the study procedure or fatigue. Method effects therefore cannot be fully separated from block position.

Third, the study compared one implementation of each correction approach rather than isolating the correction method from all system-level differences. Manual correction was represented by MyFitnessPal, while collaborative correction used the research prototype with DietAI24. Database coverage, search behavior, interface design, model behavior, and system latency may therefore contribute to the observed differences. The collaborative mechanisms were also evaluated together rather than independently, so the study cannot determine how much Text/Voice Feedback, Quick Clarification, or Visualized Portion Size individually contributed to workload, efficiency, or user experience. Component and hybrid studies could examine when each mechanism is most useful and when direct editing should remain available.

Fourth, the reference data and outcome measures have limitations. SNAPMe is linked to ASA24 self-reported dietary records, which may themselves contain errors in food identity or amount \cite{larke2023snapme}; they provide an independent dietary reference but not weighed ground-truth intake. Nutrient accuracy was evaluated across ten selected photographs, and nonsignificant differences in protein, carbohydrate, and fat should not be interpreted as evidence of statistical equivalence. Nutrient agreement also does not establish semantic record correctness or the absence of newly introduced errors. Participant identifiers were unavailable in the nutrient dataset, so the photo-level tests could not account for repeated observations within participants. The available aggregate behavioral records did not document timing boundaries or click categories and could not be linked to survey responses or image assignments; these behavioral comparisons should therefore be interpreted as exploratory. In addition, confidence, perceived control, trust, intended use, and feature ratings are subjective measures, and most were assessed with single items. Favorable confidence or trust therefore does not itself establish record accuracy or appropriate reliance.

Finally, the study was conducted with a newly developed prototype in a researcher-guided setting rather than during sustained everyday use. Participants may respond differently when correcting their own meals repeatedly over days or weeks, when speech input can be used more naturally, or when no researcher is present. The present study therefore establishes evidence about controlled correction tasks, not long-term adherence, clinical benefit, or continued use. Longitudinal studies should examine whether the lower correction effort observed in this controlled setting extends to sustained dietary logging and whether users remain able to detect and recover from incorrect AI-generated revisions over time.

\section{Conclusion}

Image-based dietary assessment can reduce the effort of creating an initial dietary record, but users still need efficient ways to correct information that is inaccurate, ambiguous, or unavailable from the image. We introduced Human--AI Collaborative Correction, which reframes correction from manual database search and field editing as \emph{communicate intent $\rightarrow$ regenerate $\rightarrow$ verify}. In a within-participant study with 30 participants, collaborative correction was associated with lower perceived workload, recorded usage time, and recorded clicks and received more favorable ratings of ease, confidence, perceived control, trust, intended use, and overall preference. The collaborative workflow also showed lower energy MAE, while observed macronutrient errors remained in a similar range to the representative commercial manual workflow. Together, the findings support intent-driven regeneration as a promising interaction pattern for reducing the repeated work involved in correcting AI-generated structured records while retaining user verification of the resulting record. In dietary assessment, lowering this correction burden may reduce barriers to initiating and sustaining food tracking, potentially making image-based dietary assessment more accessible for everyday use and supporting more consistent dietary self-monitoring over time.

\bibliographystyle{ACM-Reference-Format}
\bibliography{references}

\clearpage
\appendix
\section{Prompts Used by the MLLM Audit Agents and Visual Portion Cue Generator}
\label{app:prompts}

The following system-level prompt templates are reproduced in full. Bracketed fields denote runtime inputs. The four GPT-5.4 audit agents rank plausible candidate corrections for user selection through Quick Clarification; they do not autonomously modify the dietary record. GPT-Image-2 uses the fifth template to generate cues for Visualized Portion Size from FNDDS food and portion descriptions. User-selected corrections become context for DietAI24's subsequent estimation pass, as described in Section~\ref{sec:prompt-workflow}.

\subsection{Ingredient Confirmation Agent}
\label{app:prompt-ingredient-confirmation}

\begin{lstlisting}[style=systemprompt]
You are an ingredient confirmation agent for dietary assessment.

Input:
[MEAL_IMAGE OR MEAL_TEXT]
[CURRENT_FOOD_RECOGNITION_AND_PORTION_ESTIMATION_RESULTS]
[CURRENT_FOOD_DESCRIPTION]

Review [CURRENT_FOOD_DESCRIPTION] in the context of the meal input and the current dietary assessment.

Answer the question:

“Do you want to change [CURRENT_FOOD_DESCRIPTION]?”

Provide plausible alternative food descriptions that could replace [CURRENT_FOOD_DESCRIPTION]. Rank the candidate replacements from most to least likely for user selection.
\end{lstlisting}

\subsection{Missing Ingredient Agent}
\label{app:prompt-missing-ingredient}

\begin{lstlisting}[style=systemprompt]
You are a missing ingredient audit agent for dietary assessment.

Input:
[MEAL_IMAGE OR MEAL_TEXT]
[CURRENT_FOOD_RECOGNITION_AND_PORTION_ESTIMATION_RESULTS]

Review the meal input together with the current recognized food items.

Answer the question:

“Do you want to add any ingredients?”

Identify plausible food items that may be missing from the current dietary assessment and provide them as candidate additions for user selection. Rank the candidates from most to least likely.
\end{lstlisting}

\subsection{Added Fat, Sauce, or Seasoning Agent}
\label{app:prompt-added-fat}

\begin{lstlisting}[style=systemprompt]
You are an added fat, sauce, and seasoning audit agent for dietary assessment.

Input:
[MEAL_IMAGE OR MEAL_TEXT]
[CURRENT_FOOD_RECOGNITION_AND_PORTION_ESTIMATION_RESULTS]

Review the meal input together with the current dietary assessment.

Answer the question:

“Do you want to change any added fats, sauces, or seasonings?”

Identify plausible added fats, sauces, dressings, condiments, or seasonings associated with the meal and provide them as candidate corrections for user selection. Rank the candidates from most to least likely.
\end{lstlisting}

\subsection{Portion Review Agent}
\label{app:prompt-portion-review}

\begin{lstlisting}[style=systemprompt]
You are a portion review agent for dietary assessment.

Input:
[MEAL_IMAGE OR MEAL_TEXT]
[CURRENT_FOOD_RECOGNITION_AND_PORTION_ESTIMATION_RESULTS]
[CURRENT_FOOD_DESCRIPTION]

Review the estimated portion of [CURRENT_FOOD_DESCRIPTION] in the context of the meal input and current dietary assessment.

Answer the question:

“Do you want to change the portion of [CURRENT_FOOD_DESCRIPTION]?”

Provide plausible alternative portion sizes for [CURRENT_FOOD_DESCRIPTION] as candidate corrections for user selection. Rank the candidate portions from most to least likely.
\end{lstlisting}

\clearpage
\subsection{Visual Portion Cue Generation Prompt}
\label{app:prompt-visual-portion}

\begin{lstlisting}[style=systemprompt]
You are a food portion visualization model.

Input:
[FNDDS_FOOD_CODE_DESCRIPTION]
[FNDDS_PORTION_SIZE_DESCRIPTION]

Generate a realistic visual portion cue depicting the food specified by [FNDDS_FOOD_CODE_DESCRIPTION] in the amount represented by [FNDDS_PORTION_SIZE_DESCRIPTION].

Accurately represent both the food described by [FNDDS_FOOD_CODE_DESCRIPTION] and the portion specified by [FNDDS_PORTION_SIZE_DESCRIPTION]. If the portion description specifies a serving form, measurement unit, or container, visually represent the corresponding serving form and amount.

Show the food portion clearly using an appropriate serving container, a consistent viewing angle, neutral lighting, a plain background, and a centered composition. The generated image will be used as a visual reference to support portion-size selection in the dietary-assessment interface.
\end{lstlisting}

\clearpage
\section{Measurement and Analysis Details}
\label{app:measurement-analysis}

This appendix describes survey scoring, aggregate behavioral measures, nutrient-error calculations, and statistical analyses.

\subsection{Survey Scoring and Interpretation}

The adapted Raw TLX uses the unweighted scoring approach described by Hart \cite{hart2006tlx}. Composite and dimension scores are reported on a 0--100 scale, with higher values indicating greater burden. The unweighted composite requires complete responses to all six dimensions. Performance is oriented from Perfect to Failure and measures perceived task performance, not record accuracy. Ease, confidence, perceived control, trust, and intended everyday use use 1--5 responses, with higher ratings indicating a more favorable experience. Ease ranges from very difficult to very easy, confidence from not at all to extremely confident, and control, trust, and intended use use agreement scales. The export contains one response per participant per method, even where the questionnaire labels items as post-task.

Manual-condition feature items address food search, portion entry, and the manageability of correction steps. Collaborative-condition items cover expected behavior, Quick Clarification relevance, usefulness, option availability and list manageability, and Visualized Portion Size clarity and usefulness. Helpfulness wording permits experienced or anticipated usefulness, so favorable ratings do not establish feature use.

\subsection{Exploratory Aggregate Behavioral Analysis}
\label{app:behavioral-analysis}

The behavioral extract contained paired condition records for usage time and food clicks. Following the study context, its N and M codes were interpreted as collaborative and manual correction, respectively. Recorded minute-and-second strings were converted to seconds, and total clicks were summed across the five food columns within each participant and method. All available pairs and recorded zero counts were retained without imputation or outlier removal. Zero clicks do not establish successful correction or feature non-use. The extract does not document timing boundaries, gesture categories, task order, or image assignments; method order is established by the protocol as manual first and collaborative second, and its participant identifiers cannot be linked to the survey export in the available records.

Two-sided Wilcoxon signed-rank tests compared paired condition totals using the asymptotic method, zero-difference omission, and no continuity correction. Holm adjustment covered the two aggregate outcomes as an exploratory analysis choice. Differences and rank-biserial effects are collaborative minus manual, with average ranks for tied absolute differences; negative values indicate lower collaborative totals. Pointwise percentile 95\% confidence intervals for mean and median paired differences and rank-biserial effects used 20,000 participant-pair bootstrap resamples, recomputing each estimate per resample (seed 20260910 for time and 20260911 for clicks). Undefined bootstrap effects were omitted. The click-effect interval of [$-1$, $-1$] reflects the common direction of all observed pairs and does not imply population certainty. Percentage reductions compare condition means, calculated as $100(1-\bar{x}_{\mathrm{Collaborative}}/\bar{x}_{\mathrm{Manual}})$; they are not averages of participant-specific percentage reductions. Figure~\ref{fig:combined-results}(a) displays means and one sample SD across participants. These aggregate comparisons cannot separate method from fixed order and do not adjust for image assignment.

\subsection{Exploratory Nutrient Accuracy Analysis}
\label{app:nutrient-analysis}

The analysis retained all available entries, with no outlier removal. Absolute errors in energy, protein, carbohydrate, and fat were calculated against the ASA24/SNAPMe reference values, averaged within each photo and method, and then equally across ten photos. The RQ3 panel of Figure~\ref{fig:combined-results} reports these means and the sample SD across photo-level means.

Exact two-sided paired sign-flip tests enumerated all $2^{10}=1{,}024$ sign assignments to the collaborative-minus-manual differences. Each raw $p$ value is the proportion of assignments with an absolute mean difference at least as large as observed, including ties. Holm adjustment covered the four outcomes. Validity of this enumeration requires the joint null distribution to be invariant to the applied sign flips, as with independent symmetric photo-level differences \cite{winkler2014permutation}. These exploratory tests use photos as the pairing unit; participant-level dependence could not be assessed without participant identifiers, and method differences remain confounded with fixed order. The tests describe differences between the observed workflows rather than isolating a causal effect of correction method.

\subsection{Paired Survey Analysis}

Respondents had complete paired numeric survey ratings; the survey analysis retained every pair without imputation or exclusions. We calculated median paired differences, two-sided Wilcoxon signed-rank tests \cite{wilcoxon1945ranking}, and matched-pairs rank-biserial effects \cite{kerby2014difference}. Differences are collaborative minus manual; positive differences and effects denote higher collaborative ratings.

Median paired differences and rank-biserial effects receive percentile 95\% confidence intervals from 20,000 participant-level bootstrap resamples, preserving each participant's condition pair \cite{efron1994bootstrap}. Intervals are pointwise, not multiplicity-adjusted; boundary intervals reflect the observed resampling support, not population certainty.

The Raw TLX composite retains its unadjusted test. Holm adjustment \cite{holm1979multiple} applies separately to the six workload dimensions and to ease, confidence, control, trust, and intended use. These comparison families are retained from the existing analysis, without ranking the three RQ domains by importance. Preference, perceived easier method, and method-specific feature ratings are descriptive. Survey responses were not linked to image-set allocation metadata. Fixed method order prevents separate estimation of method and order effects. Preference intervals use pointwise Wilson 95\% confidence intervals \cite{wilson1927inference} with all responses, including those reporting no difference, in the denominator. Workload and experience plots show means and one-sample SD, complementing the median and quartile tables.

\end{document}